\documentclass[preprint,12pt]{elsarticle}

\usepackage{amssymb}
\usepackage{url}
\usepackage{hyperref}

\usepackage{microtype}

\usepackage[T1]{fontenc}

\usepackage{lineno}

\journal{Nuclear Physics A}

\newcommand{\qhat}{\hat{q}}
\newcommand{\trento}{T$\mathrel{\protect\raisebox{-2.1pt}{R}}$ENTo}
\newcommand{\td}{\mathrm{d}}
\newcommand{\cE}{\mathcal{E}}
\newcommand{\TT}[1]{\mathrm{#1}}
\newcommand{\As}{\alpha_{\mathrm{s}}}
\newcommand{\Raa}{\ensuremath{R_{\mathrm{AA}}}}
\newcommand{\sqrts}{\ensuremath{\sqrt{s_{\mathrm{NN}}}}}

\begin{document}

\begin{frontmatter}



\title{Predictions for the sPHENIX physics program}


\affiliation[uncg]{organization={University of North Carolina, Greensboro}}
\affiliation[cern]{organization={CERN}}
\affiliation[mit]{organization={Massachusetts Institute of Technology}}
\affiliation[subatech]{organization={SUBATECH (IMT Atlantique, Universit\'{e} de Nantes, IN2P3/CNRS)}}
\affiliation[gsu]{organization={Georgia State University}}
\affiliation[belgrade]{organization={Institute of Physics Belgrade}}
\affiliation[berkeley]{organization={University of California Berkeley}}
\affiliation[santiago]{organization={Universidade de Santiago de Compostela}}
\affiliation[barcelona]{organization={Universitat de Barcelona}}
\affiliation[heidelberg]{organization={Institut f\"ur Theoretische Physik, Universit\"at Heidelberg}}
\affiliation[bnl]{organization={Brookhaven National Laboratory}}
\affiliation[rbrc]{organization={RIKEN BNL Research Center}}
\affiliation[cnrs]{organization={Ecole Polytechnique}}
\affiliation[lanl]{organization={Los Alamos National Laboratory}}
\affiliation[ucla]{organization={University of California Los Angeles}}
\affiliation[wayne]{organization={Wayne State University}}
\affiliation[infn]{organization={INFN, Sezione di Torino}}
\affiliation[oviedo]{organization={Departamento de F\'{i}sica, Universidad de Oviedo}}
\affiliation[ictea]{organization={Instituto Universitario de Ciencias y Tecnolog\'{i}as Espaciales de Asturias (ICTEA)}}
\affiliation[cu]{organization={University of Colorado Boulder}}
\affiliation[uiuc]{organization={University of Illinois Urbana Champaign}}
\affiliation[ksu]{organization={Kent State University}}
\affiliation[bergen]{organization={Bergen University}}
\affiliation[ccnu]{organization={Central China Normal University}}
\affiliation[lbnl]{organization={Lawrence Berkeley National Laboratory}}

\fntext[fn1]{Workshop organizers}
\cortext[cor1]{Corresponding author, \url{dvp@bnl.gov}}
\fntext[jetscape]{Contributions on behalf of the JETSCAPE Collaboration}


\author[uncg]{Ron Belmont}
\author[cern]{Jasmine Brewer}
\author[mit]{Quinn Brodsky}
\author[subatech]{Paul Caucal}
\author[gsu]{Megan Connors\fnref{fn1}}
\author[belgrade]{Magdalena Djordjevic}
\author[berkeley,lbnl]{Raymond Ehlers\fnref{jetscape}}
\author[santiago,barcelona]{Miguel A. Escobedo}
\author[santiago]{Elena G. Ferreiro}
\author[heidelberg]{Giuliano Giacalone}
\author[bnl,rbrc]{Yoshitaka Hatta}
\author[cnrs]{Jack Holguin}
\author[lanl]{Weiyao Ke}
\author[ucla]{Zhong-Bo Kang}
\author[wayne]{Amit Kumar\fnref{jetscape}}
\author[cern,heidelberg]{Aleksas Mazeliauskas}
\author[bnl,rbrc]{Yacine Mehtar-Tani\fnref{fn1}}
\author[rbrc]{Genki Nukazuka\fnref{fn1}}
\author[infn,oviedo,ictea]{Daniel Pablos}
\author[cu]{Dennis V. Perepelitsa\fnref{fn1}\corref{cor1}}
\author[mit]{Krishna Rajagopal}
\author[uiuc]{Anne M. Sickles\fnref{fn1}}
\author[ksu]{Michael Strickland}
\author[bergen]{Konrad Tywoniuk}
\author[lanl]{Ivan Vitev}
\author[lbnl]{Xin-Nian Wang}
\author[ccnu]{Zhong Yang}
\author[ucla]{Fanyi Zhao}

\begin{abstract}
sPHENIX is a next-generation detector experiment at the Relativistic Heavy Ion Collider, designed for a broad set of jet and heavy-flavor probes of the Quark-Gluon Plasma created in heavy ion collisions. In anticipation of the commissioning and first data-taking of the detector in 2023, a RIKEN-BNL Research Center (RBRC) workshop was organized to collect theoretical input and identify compelling aspects of the physics program. This paper compiles theoretical predictions from the workshop participants for jet quenching, heavy flavor and quarkonia, cold QCD, and bulk physics measurements at sPHENIX.
\end{abstract}


\end{frontmatter}


\tableofcontents

\section{Introduction}
\label{}

Relativistic collisions of heavy ion nuclear beams produce Quark-Gluon Plasma (QGP), a high-density and high-temperature phase of matter comprised of deconfined and strongly-interacting quarks and gluons~\cite{Elfner:2022iae,Busza:2018rrf}. The QGP is the primordial substance which dominated the observable universe in the microseconds after its creation in the Big Bang, and has been the subject of intensive experimental studies for decades at facilities around the world. In the modern era of collider-based experiments, QGP has been produced and studied at the Relativistic Heavy Ion Collider (RHIC) at Brookhaven National Laboratory (BNL) since 2000 and at the Large Hadron Collider (LHC) at CERN since 2010.

The QGP exhibits several remarkable many-body phenomena, such as a collective expansion which is describable by relativistic hydrodynamics with a near-perfect fluidity~\cite{Heinz:2013th}, first observed by experiments at RHIC~\cite{BRAHMS:2004adc,PHOBOS:2004zne,STAR:2005gfr,PHENIX:2004vcz} and subsequently confirmed by experiments at the LHC~\cite{ALICE:2010suc,ATLAS:2011ah,CMS:2012zex}. 
The QGP is composed of quarks and gluons whose interactions are described by quantum chromo-dynamics (QCD), the theory of the strong nuclear interaction. Despite this, it is not understood how the properties and observed long-wavelength behavior of the QGP emerge from these fundamental degrees of freedom - a key open question in nuclear physics~\cite{Aprahamian:2015qub}. 
To address this question, high transverse momentum ($p_\mathrm{T}$) jets or heavy-flavor hadrons that are formed from the fragmentation of hard partons produced in the early stages of the nucleus–nucleus collision, have been recognized as unique probes of the QGP over a wide range of scales.  
As they propagate through the expanding, cooling medium, their interactions with the QGP probe its properties over momentum scales ranging from the deeply perturbative to those comparable to fluid scales. Experimental measurements of the ``quenching'' of high-$p_\mathrm{T}$ jets and the modification of heavy-flavor hadrons have been used for this purpose at RHIC, and then greatly expanded under the later-generation experiments at the LHC~\cite{Cunqueiro:2021wls,Apolinario:2022vzg}.

To definitively address the above open questions and complete the scientific mission of RHIC, the sPHENIX experiment was  designed as a new, next-generation collider detector to measure jet and heavy-flavor observables with a level of precision not previously achievable at RHIC. The particular timeliness and necessity of the sPHENIX physics program has been widely recognized, for example in white papers contributed by the community to the U.S. Long-Range Plan for Nuclear Science process starting in 2022~\cite{Arslandok:2023utm,Achenbach:2023pba}. As this paper is being written, sPHENIX is undergoing final installation in the experimental hall of the former PHENIX detector, with first collisions for detector commissioning expected in May 2023. 

In order to continue to develop the theoretical context for the first sPHENIX data and strengthen the scientific relationships between the sPHENIX Collaboration and the theoretical community, a three-day workshop sponsored by the RIKEN BNL Research Center (RBRC) was held at BNL in July 2022 with over one hundred registered participants, entitled ``Predictions for sPHENIX''~\cite{workshop}. The workshop was held in a plenary style, with heavy-ion theorists asked to highlight interesting physics which is potentially accessible with sPHENIX and, when possible, to give concrete predictions before the arrival of first data. This paper represents a summary of the predictions and physics discussion at the workshop, with written contributions from the theory speakers and their close collaborators collected in the sub-sections that follow. The intention of the authors is that the summaries below help motivate further theoretical work which is focused on the quantitative extraction of physics information from the future sPHENIX data.

\section{sPHENIX Experiment at RHIC}

\begin{figure}
  \centering
    \includegraphics[width=0.90\textwidth]{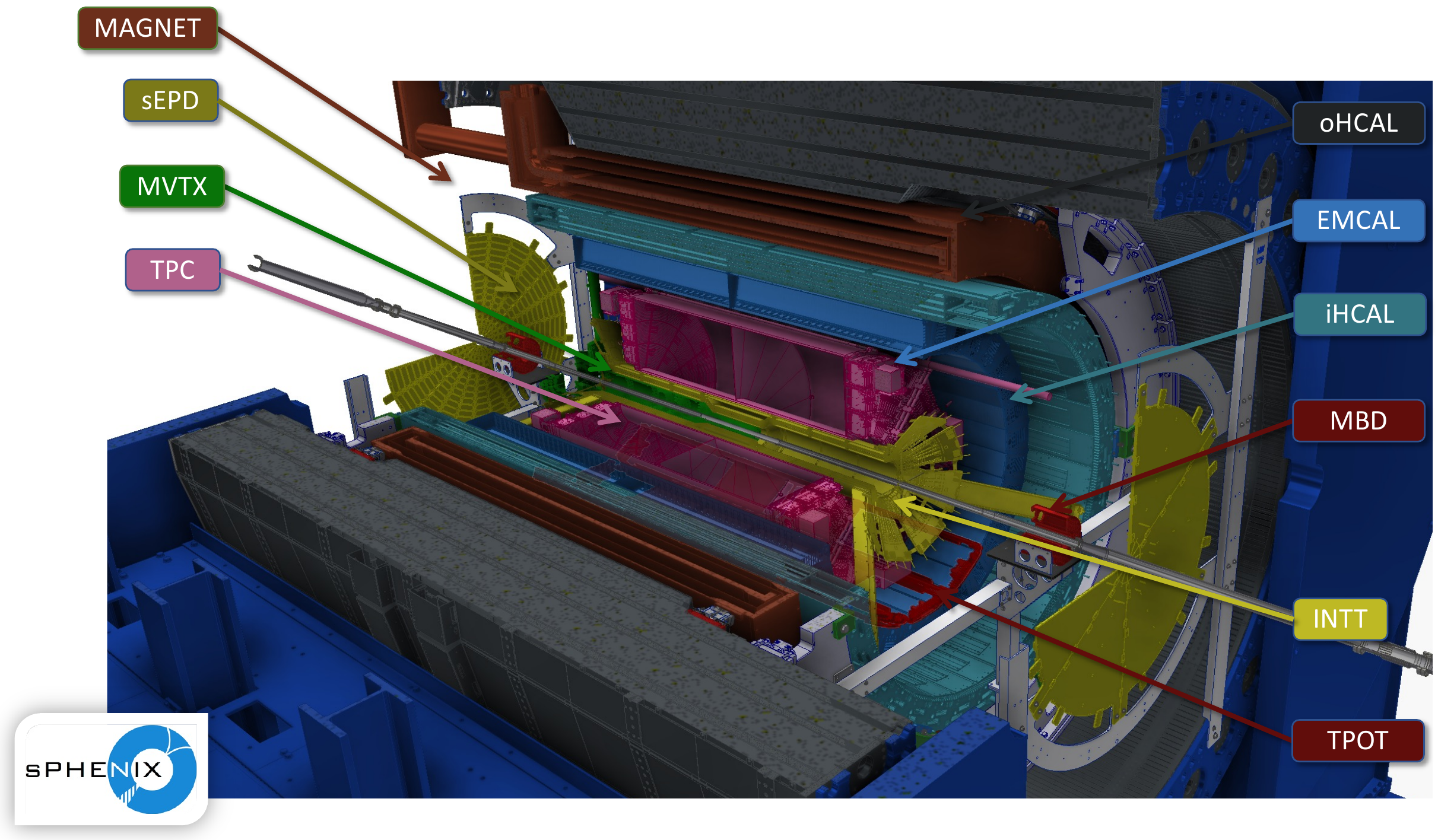}
    \caption{3-D rendering of the sPHENIX detector. An azimuthal slice has been removed to better show the component sub-detectors. Not shown are the ZDC and SMD, which are situated out of frame on either side of the collision point.}\label{Fig:sPHENIX}
\end{figure}

The sPHENIX experiment is a new collider detector at RHIC, situated at the 8 o'clock position of the ring, in the interaction region of the predecessor PHENIX experiment. Some principal highlights of sPHENIX relevant to the physics program include its high data rate, large acceptance, precision tracking close to the vertex, high momentum resolution, hadronic calorimetry, unbiased triggering in $p$+$p$ collisions, and streaming readout capability of the tracking detectors. The detector and its major subsystems are rendered in Fig.~\ref{Fig:sPHENIX}.

The sPHENIX detector is constructed around and within a superconducting solenoid magnet, originally used in the BaBar experiment~\cite{BaBar:2001yhh}, which produces an axial magnetic field that reaches 1.4~Tesla in its center. The tracking system consists of four detectors, listed here in order of their proximity to the beampipe. The monolithic active pixel sensor (MAPS) vertex detector (MVTX) consists of three layers providing precision position information to identify the originating vertex of charged-particle tracks. An intermediate silicon tracker (INTT) consists of four layers which provide two hit measurements. The INTT has a fast timing resolution and is used to connect the track trajectories between the MVTX and Time Projection Chamber (TPC). The compact TPC has a gateless design and provides the large lever arm for the momentum measurement. The MVTX, INTT, and TPC trackers have full azimuthal acceptance, and an acceptance in pseudorapidity of $\left|\eta\right| < 1.1$ for events with a collision vertex within $\left|z\right| < 10$~cm. Finally, the TPC Outer Tracker (TPOT) detector consists of eight rectangular panels situated below the TPC which provide a confirmation hit to aid in the determination of corrections due to TPC space charge distortions. All the tracking detectors are read out in a continuous streaming mode. Some aspects of the tracking reconstruction in sPHENIX are discussed in Ref.~\cite{Osborn:2021zlr}.

The calorimeter system comprises the electro-magnetic calorimeter (EMCal) and a hadronic calorimeter split into an Inner (IHCal) and Outer (OHCal) section on either side of sPHENIX solenoid magnet. The hermetic system covers the full acceptance in azimuth and $\left|\eta\right| < 1.1$. The EMCal is composed of towers filled with tungsten and scintillating fiber with a 2-D projective design. The segmentation of the towers is $\Delta\eta\times\Delta\phi \approx 0.025\times0.025$, and the EMCal provides approximately twenty radiation lengths for high-resolution measurements of the energy deposited by photons and electrons. The hadronic calorimeter is composed of aluminum (IHCal) or steel (OHCal) plates interleaved with scintillating tiles. The IHCal is particularly important for catching the start of hadronic showers before the inactive material in the magnet. The IHCal and OHCal together provide approximately five nuclear interaction lengths, with a segmentation of $\Delta\eta\times\Delta\phi \approx 0.1\times0.1$, and good energy resolution appropriate for calorimeter-based measurements of high-$p_\mathrm{T}$ hadrons and jets. The OHCal also doubles as the flux return for the magnet. Prototypes of the calorimeter system, and their description in simulation, have been characterized using test beam at the Fermilab Test Beam Facility~\cite{sPHENIX:2017lqb,Aidala:2020toz}. 

The tracking and calorimeter barrel detectors are complemented with a suite of forward detectors for minimum-bias triggering and event categorization. These include the Minimum-Bias Detector (MBD), the sPHENIX Event Plane Detector (sEPD), and the Zero-Degree Calorimeter (ZDC) which includes a Shower Max Detector (SMD). The MBD consists of quartz radiator tubes situated on either side of the detector at $3.51 < \left|\eta\right| < 4.61$, and will be used for minimum-bias triggering and as a possible detector for global centrality determination. The sEPD consists of two wheels of finely-segmented scintillating plastic tiles situated at $2.0 < \left|\eta\right| < 4.9$, and is intended for event plane determination. Two ZDC detectors are situated 18~m away on either side of the collision point, which measure primarily neutrons from the fragmenting nuclei and aid in the event selection. Finally, the SMD consists of layers of scintillating tiles situated behind the first ZDC module, providing spatial information. It may be used to determine the first-order event plane ($\Psi_1$) in Au+Au data-taking and as a polarimeter in polarized $p$+$p$ and $p$+Au data-taking. 

The sPHENIX scientific collaboration has proposed a three-year data-taking program which would accomplish the scientific mission of the experiment. That plan and some examples of the statistical and kinematic reach for specific physics measurements are detailed in the sPHENIX Beam Use Proposal~\cite{sPHENIXBUP} and briefly summarized at the start of each section below. The proposed running scenario begins with Au+Au running in 2023 to commission the detector and provide first physics data, transversely polatized $p$+$p$ and $p$+Au running in 2024 to provide the reference data for the Au+Au program and for cold QCD physics, and high-statistics Au+Au running in 2025 to provide an archival dataset for QGP physics with unprecedented statistical precision at RHIC. Furthermore, the sPHENIX collaboration stands ready to capitalize on additional opportunities for data-taking to deliver  physics beyond the core program, such as the detailed exploration of intermediate collision species at RHIC~\cite{Brewer:2021kiv}.

\section{Jet probes of the QGP}

A major scientific motivation for the sPHENIX experiment is its broad program of reconstructed-jet physics measurements. The large acceptance, hermetic hadronic calorimetry, high-efficiency tracking, and large data-taking rate of the detector will result in an enormous data sample of high-$p_\mathrm{T}$ reconstructed jets. Measurements of the jet kinematics can be performed using information from just the calorimeter system, or by further incorporating information from the trackers. The luminosity during the first three years of data-taking is projected to provide one million jets above $30$~GeV and fifty thousand direct photons above $20$~GeV in the most central 0--10\% Au+Au events. These capabilities will enable precision measurements of overall jet production and intra-event correlations (such as jet $v_n$, di-jet asymmetries, photon-tagged jets), the modification of the internal jet (sub-)structure, and correlated presence of particles at large angles from the jet axis. The kinematic coverage will overlap with those of the LHC at the high-$p_\mathrm{T}$ (for example, projections indicate a measurement of the jet $R_\mathrm{AA}$ out to 70~GeV) while at the same, due to the smaller underlying event, extending into a lower-$p_\mathrm{T}$ regime not previously accessible. Given the large potential of this aspect of the sPHENIX physics program, a significant focus of the workshop was dedicated towards predictions for jet probes of the QGP, which are summarized below.

\subsection{Baseline for jet and hadron nuclear modification factors in sPHENIX}


One of the most robust signals of a hot QCD medium created in nuclear collisions is the suppression
of high momentum spectra of single inclusive hadrons and jets. The commonly used experimental observable
for quantifying these energy loss phenomena is the ratio of hadron $(h)$ or jet $(j)$ spectra in AA and $pp$ collisions, which is
known as the nuclear modification factor $R^{j,h}_\mathrm{AA}$. However even in the absence of a hot QCD medium, $R^{j,h}_\mathrm{AA}$
deviates from unity due to the different partonic compositions of a proton and a nucleus. These parton distribution functions (PDFs)
are extracted from global fits of perturbative QCD calculations to a large set of experimental data. With the increasing order of perturbative 
calculations and the increasing amount of experimental data, both proton and nuclear PDFs are constantly being improved.
Nevertheless, uncertainties in nuclear PDFs (nPDFs) remain the dominant uncertainty in the theoretical calculations of $R^{j,h}_\mathrm{AA}$ baseline in the absence of hot QCD medium~\cite{Huss:2020dwe,Brewer:2021tyv}.

Here the theoretical baseline of jet and hadron  $R_\mathrm{AA}$ at $\sqrt{s}=200\,\mathrm{GeV}$ is computed in minimum bias oxygen-oxygen (OO) and gold-gold (AuAu) collisions.
\begin{equation}
  \left. R_\mathrm{AA}^{h,j}\right|_\mathrm{min-bias} =  \frac{1}{A^2} \frac{d\sigma^{h,j}_\mathrm{AA}/dp_Tdy}{d\sigma^{h,j}_{pp}/dp_Tdy}
.\end{equation}
Three recent extractions of nuclear PDFs are employed: EPPS21~\cite{Eskola:2021nhw}, nNNPDF3.0~\cite{AbdulKhalek:2022fyi} and TUJU21~\cite{Helenius:2021tof} with corresponding proton PDFs.

\begin{figure*}[t]
  \centering
\includegraphics[width=0.5\linewidth]{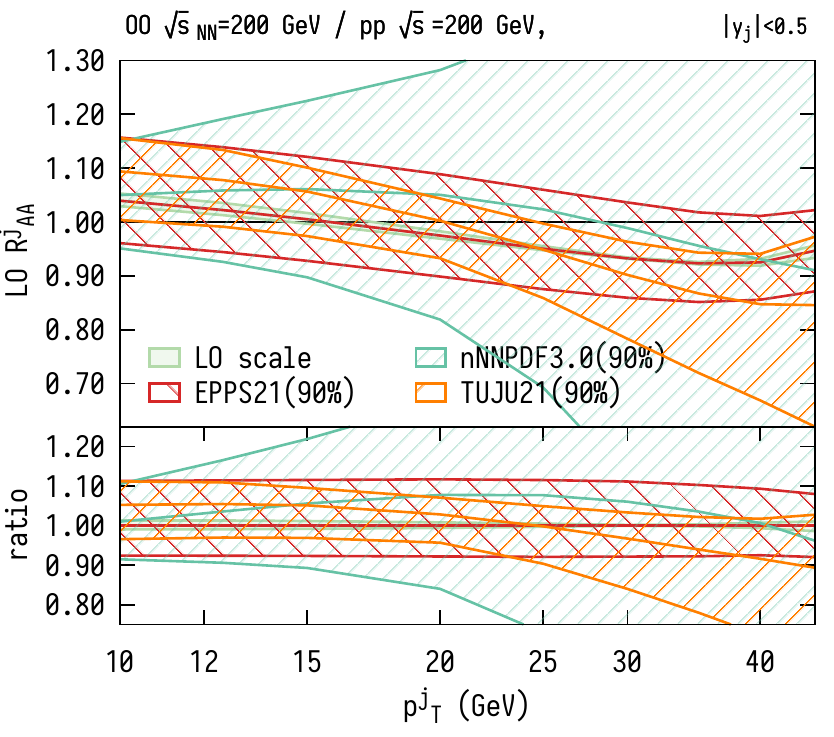}%
\includegraphics[width=0.5\linewidth]{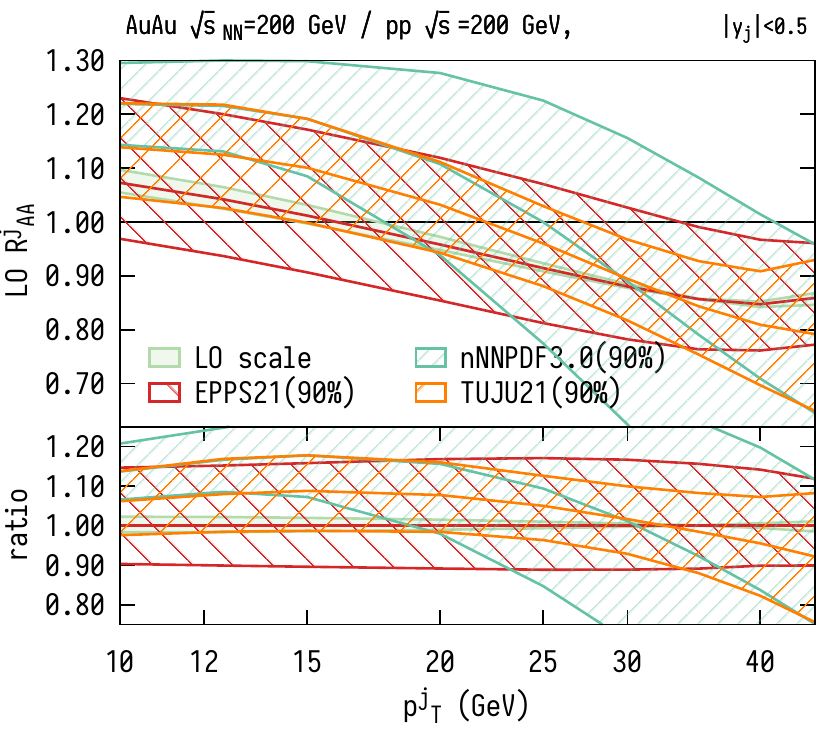}
\caption{Baseline (no energy loss) jet nuclear modification factor for (left) oxygen-oxygen and (right) gold-gold collisions. Blue solid band shows the cancellation of correlated scale uncertainties at leading order. Hatched bands show 90\% confidence intervals of combined nuclear and proton PDFs. Lower panels show $R^{j}_\mathrm{AA}$ normalized to the central EPPS21 line.}\label{fig:jetRAA_AM}
\end{figure*}

Fig.~\ref{fig:jetRAA_AM} shows the jet nuclear modification factor $R_\mathrm{AA}^j$, for which the correlated variation of factorization and renormalization scales give negligible scale uncertainty band. Although these calculations were done only at leading order for partonic jets, previous computations at next-to-leading order have shown that $R^{j}_\mathrm{AA}$ is rather insensitive to the perturbative order and showering effects~\cite{Huss:2020dwe,Brewer:2021tyv}. For AuAu collisions Fig.~\ref{fig:jetRAA_AM} (right) shows that the nPDF uncertainties for nPDF sets are of order of 10\% at $p_T^j=10\,\mathrm{GeV}$ and grow for larger momentum (especially for nNNPDF3.0). There is a slight tension between EPPS21 and nNNPDF3.0 extractions, while TUJU21 results sit in between. This indicates that even for large nuclei for which there are collider data available, different nPDF extractions could result in non-negligible differences. Finally, for $p_T^j>25\,\mathrm{GeV}$ there is a downward trend in  $R_\mathrm{AA}^j$ and therefore any conclusions about the magnitude of jet quenching in AuAu should take into account this additional suppression.
Fig.~\ref{fig:jetRAA_AM} (left) shows the analogous results for OO collisions. Surprisingly, the agreement between the central values of different nPDF sets is better for OO than AuAu. It could be in part due to the general reduction of nuclear modification for lighter nuclei. However, the nNNPDF3.0 set, which does not impose particular $A$ dependence in its extraction, results in much wider uncertainty bands with over 20\% above  $p_T^j>20\,\mathrm{GeV}$, which points to the lack of experimental data for light nuclei in nPDF extractions. If a (possibly small) jet quenching signal was to be observed in OO collisions, the uncertainties of nPDF should be reduced.

\begin{figure*}[t]
  \centering
\includegraphics[width=0.5\linewidth]{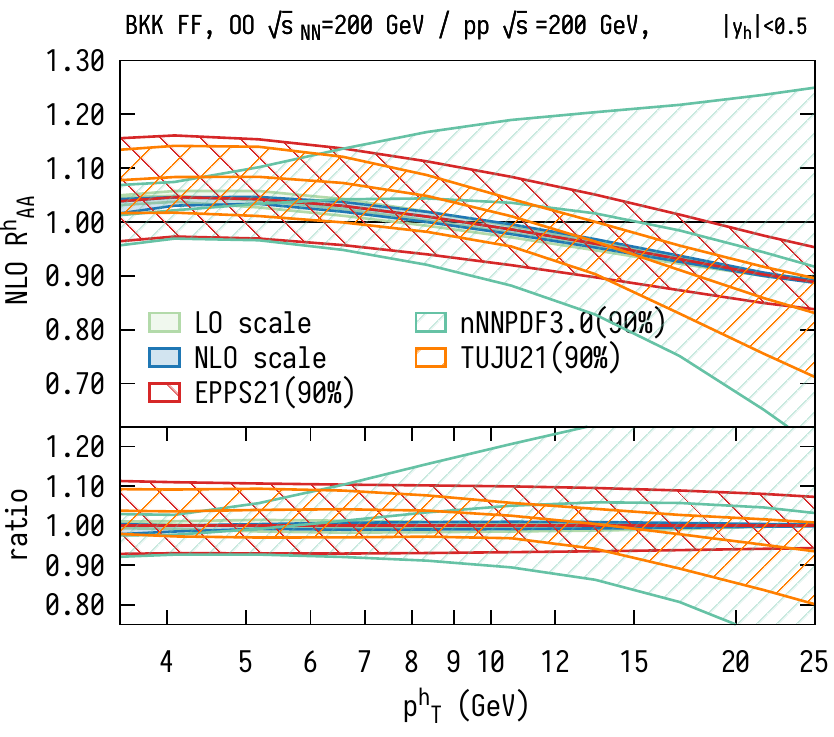}%
\includegraphics[width=0.5\linewidth]{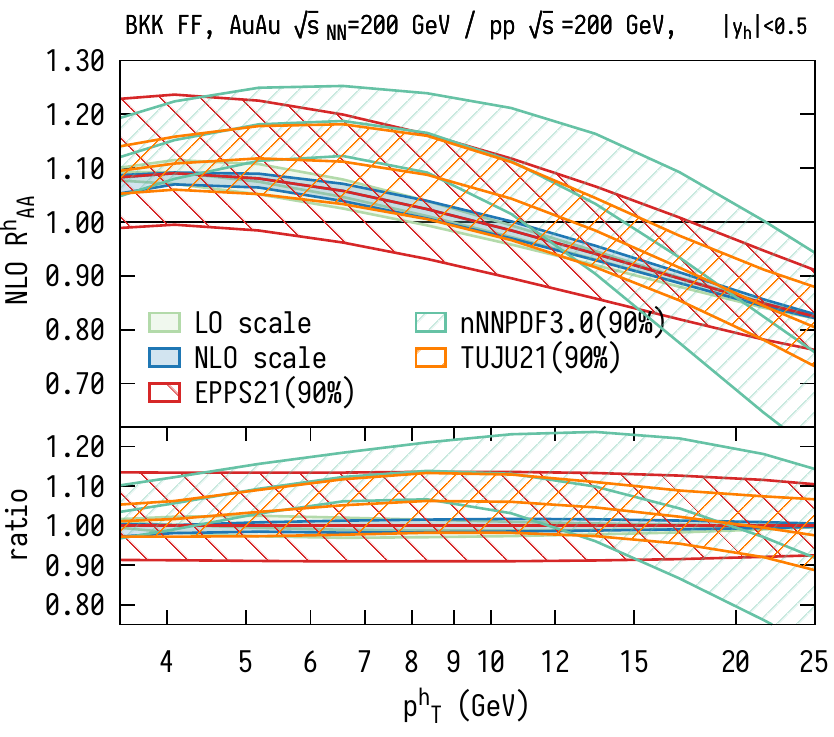}
\caption{Baseline (no energy loss) hadron nuclear modification factor for (left) oxygen-oxygen and (righ) gold-gold collisions. Green and blue solid bands show the cancellation of correlated scale uncertainties at leading and next-to-leading order. Hatched bands show 90\% confidence intervals of combined nuclear and proton PDFs. Lower panels show $R^{h}_\mathrm{AA}$ normalized to the central EPPS21 line.}\label{fig:hadronRAA_AM}
\end{figure*}

Fig.~\ref{fig:hadronRAA_AM} shows the charged hadron nuclear modification factor $R_\mathrm{AA}^h$ computed at next-to-leading order, using the INCNLO code~\cite{Aversa:1988vb}\footnote{\url{http://lapth.cnrs.fr/PHOX_FAMILY/readme_inc.html}} with LHAPDF grid support~\cite{Buckley:2014ana}.
 For the charged hadrons, the sum of BKK pion and kaon fragmentation functions (FFs)~\cite{Binnewies:1994ju} was used.
Different choices in the FFs result in negligible uncertainty compared to the nPDF uncertainty~\cite{Huss:2020dwe,Brewer:2021tyv}.
Similar to the jet  $R_\mathrm{AA}$, there is a good cancellation of scale uncertainties both at leading and next-to-leading order. For AuAu again there is a slight
tension between EPPS21 and nNNPDF3.0 results. For AuAu the uncertainties and consistency between different nPDF sets are better for hadrons than jets as the 90\% confidence intervals are contained in $\pm25\%$ bands (see lower panel of Fig.~\ref{fig:hadronRAA_AM} (right)) in the studied momentum range. For OO collisions the uncertainty bands becomes over 20\% above $p_T^h>15\,\mathrm{GeV}$.

In summary, this contribution presents the baseline calculations of jet and hadron $R_\mathrm{AA}$ in the absence of hot medium effects for OO and AuAu collisions at  $\sqrt{s}=200\,\mathrm{GeV}$. An experimentally measured deviation from these baselines would be a clear signal of additional physics to the perturbative vacuum picture of high-energy particle collisions. Although jet quenching and energy loss phenomena have been observed previously in AuAu collisions, the quantified theory uncertainties of a perturbative baseline are important in the precision quantification of hot QCD medium effects. In the case of OO collisions,
a measurement of statistically significant deviation from the computed bands in Fig.~\ref{fig:jetRAA_AM} (left) and Fig.~\ref{fig:hadronRAA_AM} (left) would signify the discovery of energy loss phenomena in small collision systems with an average of  $\sim 10$ colliding nucleons.

\subsection{JetMed predictions for sPHENIX}

JetMed is a parton shower in heavy-ion collisions based on the factorization between vacuum-like emissions (VLEs) and medium-induced radiations as described in \cite{Caucal:2018dla,Caucal:2019uvr}. This picture of jet evolution, which relies on well-controlled approximations in pQCD, turns out to be in good qualitative agreement with LHC data on high $p_t$ jets \cite{Caucal:2019uvr,Caucal:2020xad,Caucal:2020uic}. Even though the underlying approximations are less justified for the lower $p_t$ jets measured at RHIC, it would be interesting to confront this model with future sPHENIX data.  
In particular, JetMed includes color coherence effects via the in-medium phase space for VLEs which is constrained by the coherence angle $\theta_c$. Unraveling the role of color coherence in jet quenching is an active field of investigation both at RHIC and at the LHC \cite{Andrews:2018jcm,ALICE:2019ykw,STAR:2021kjt,ALargeIonColliderExperiment:2021mqf}. This $\theta_c$ angle scales like $\theta_c\sim 2/(\hat{q}L^3)^{1/2}$ for a dense medium with average quenching parameter $\qhat$ and size $L$ \cite{Mehtar-Tani:2010ebp,Casalderrey-Solana:2011ule,Casalderrey-Solana:2012evi}. Hence, since $\qhat$ at RHIC energies is typically smaller than $\qhat$ at the LHC, $\theta_c$ is larger at RHIC than at the LHC (for a similar colliding system size $L$). In principle, this should facilitate its measurement.

\begin{figure}
  \centering
    \includegraphics[width=0.48\textwidth]{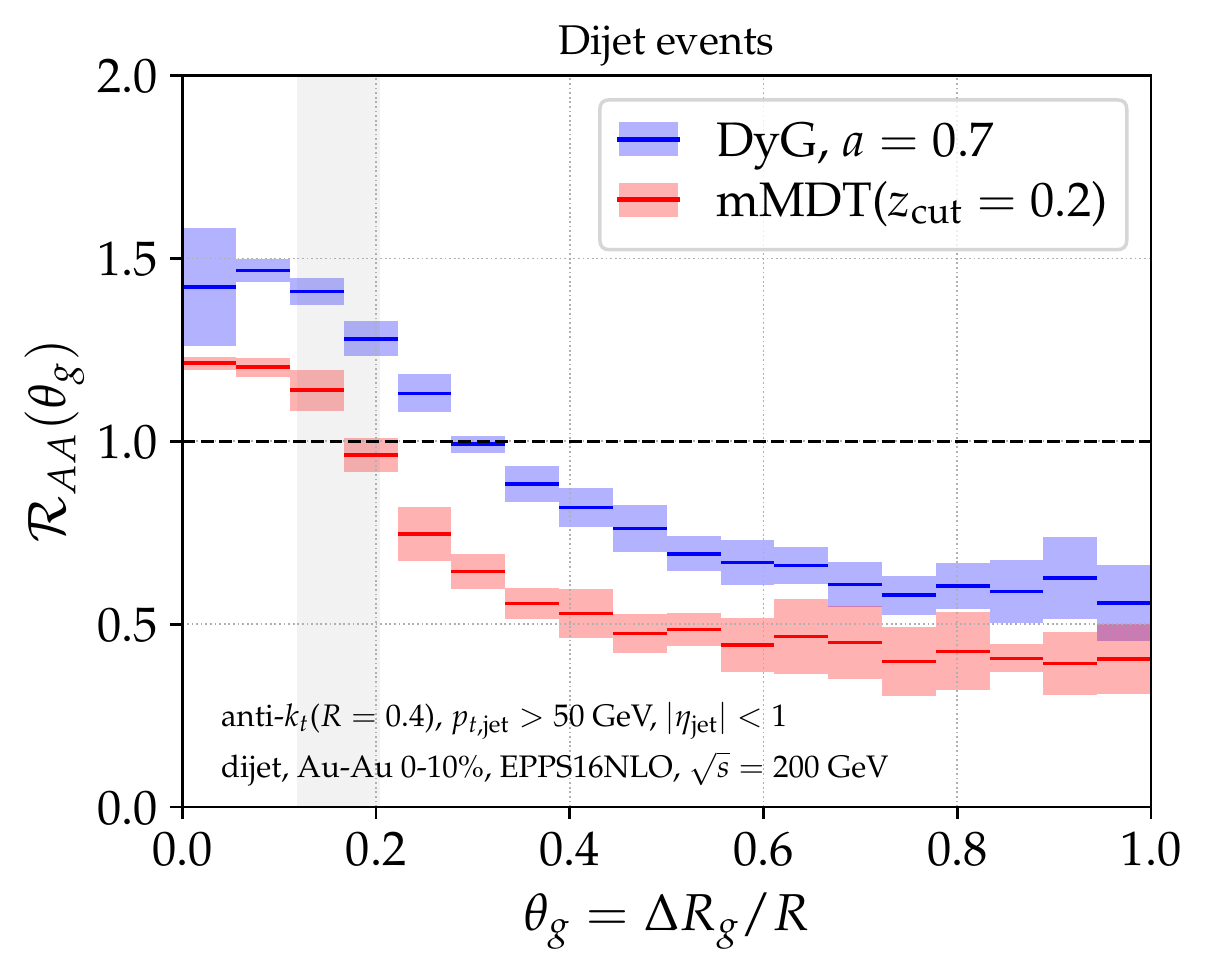}
    \includegraphics[width=0.48\textwidth]{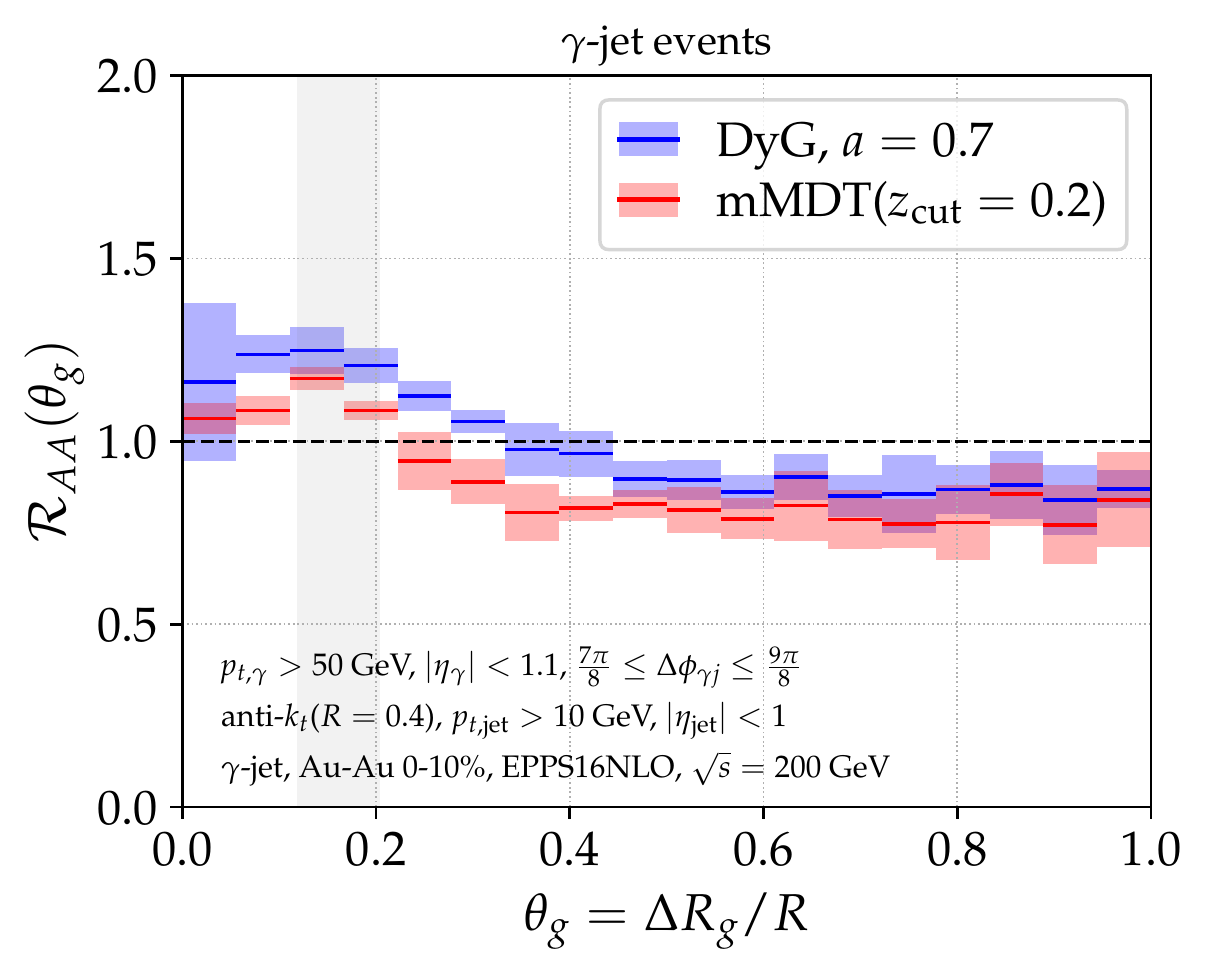}
    \caption{JetMed predictions for the $R_\mathrm{AA}$ for dijet (left) and $\gamma$-jet (right) processes in sPHENIX.}\label{Fig:jetmed-predictions}
\end{figure}

Jet substructure techniques are ideal tools to probe color (de)coherence and potentially measure $\theta_c$. Among those, the Soft Drop \cite{Larkoski:2014wba} or Dynamically groomed (DyG) \cite{Mehtar-Tani:2019rrk} jet radius have shown good sensitivity to $\theta_c$ due to a rather simple physical mechanism \cite{Caucal:2020uic,Caucal:2021cfb}. Namely, the medium acts as a filter which enhances the production of small $\theta_g\le \theta_c$ jets compared to larger $\theta_g>\theta_c$ jets, as the latter lose more energy than the former due to more resolved intrajet sources for energy loss. This is shown in the left panel of Fig.\,\ref{Fig:jetmed-predictions} which displays the nuclear modification factor for the $\theta_g$ distribution obtained from Soft Drop $(\beta=0,z_{\rm cut}=0.2)$ in red and dynamical grooming with $a=0.7$ in blue (this choice of $a$ maximizes the sensitivity to $\theta_c$ while reducing non-perturbative and background effects \cite{Caucal:2021cfb}). In this figure, the uncertainty bands correspond to variations of the unphysical cut-off parameters in the parton shower and variations of the effective $\qhat$ between $0.4$--$1.2$ GeV$^2/$fm. 
One observes a clear relative enhancement of small $\theta_g$ jets and a suppression of large $\theta_g$ jets with a transition set by the angular scale $\theta_c$ (which lies inside the vertical grey band on the figure). In comparison, $R_{AA}(\theta_g)$ in $\gamma$-jet events is shown in the right panel of Fig.\,\ref{Fig:jetmed-predictions} with the event selection performed on the photon $p_t$, following the proposal of \cite{Brewer:2021hmh} to disentangle bias effects from intrinsic modifications of the shower (besides energy loss). Interestingly, $R_{AA}(\theta_g)$ is not equal to one since JetMed also accounts for relatively hard intrajet medium-induced emissions, whose typical angular scale is also set by $\theta_c$ \cite{Casalderrey-Solana:2011ule}. Such complementary studies between dijet and $\gamma$-jet events is a promising opportunity of the sPHENIX detector.

\subsection{JETSCAPE predictions for sPHENIX}
\label{JETSCAPE_kumar}


JETSCAPE is a large-scale benchmarking framework for rigorously testing and validating  physics models describing  the dynamics of soft and hard sectors in ultra-relativistic nucleus-nucleus collisions. This section presents JETSCAPE predictions for upcoming sPHENIX measurements. The soft sector is simulated using event-by-event TRENTO \cite{Moreland:2014oya} initial conditions evolved hydrodynamically with VISHNU (2+1D) code package \cite{Shen:2014vra}. The jet evolution is carried out using a multi-stage approach within the JETSCAPE framework where the high-virtuality region of the parton shower is modeled using the MATTER \cite{Majumder:2013re} event generator and the low virtuality region is simulated using the LBT \cite{He:2015pra} event generator. The details of the full model calculation can be found in Ref.~\cite{JETSCAPE:2022jer}.

\begin{figure}[t]
\centering
\includegraphics[width=0.99\textwidth]{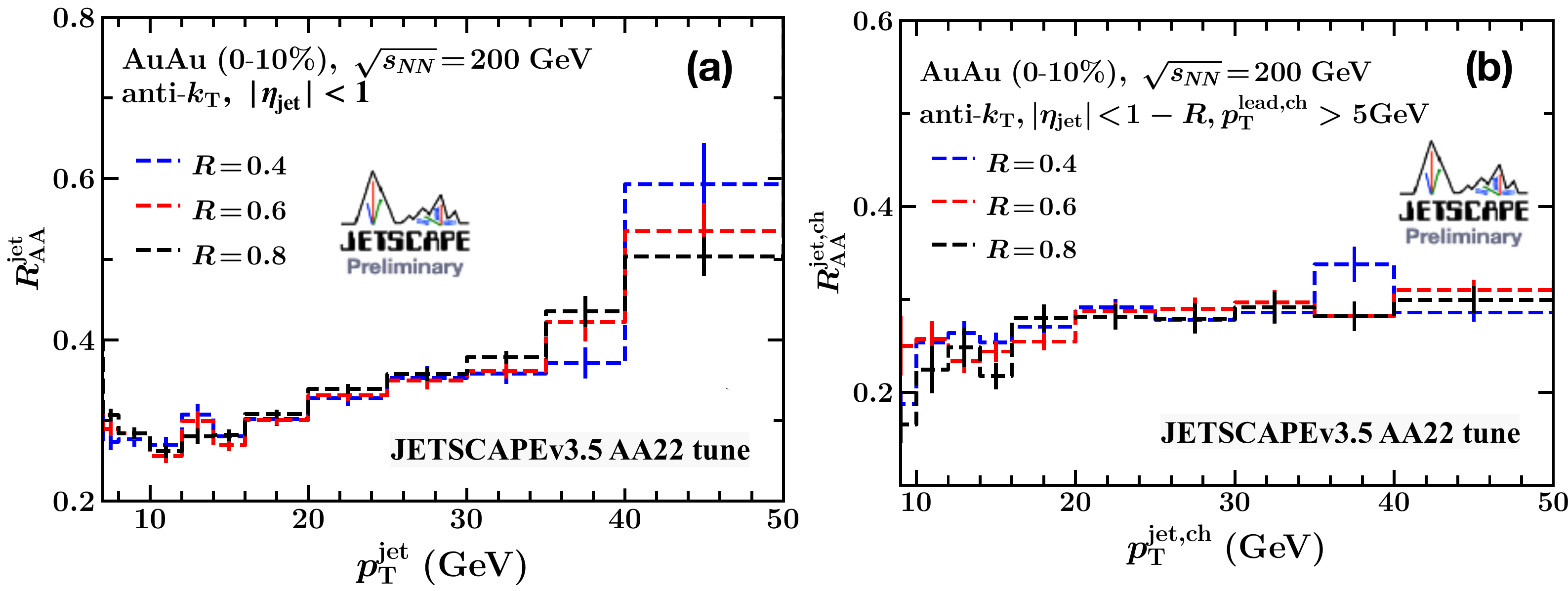}
\caption{Nuclear modification factor $R_{\mathrm{AA}}$ for full jets at most central ($0-10\%$) collisions at $\sqrt{s_{\mathrm{NN}}}=$200 GeV for jet cone sizes $R=$0.4, 0.6 and 0.8. The calculation is performed using JETSCAPEv3.5 AA22 tune described in Ref.~\cite{JETSCAPE:2022jer}. (a) Anti-k$_{\mathrm{T}}$ based inclusive jets (b) Anti-k$_{\mathrm{T}}$ based charged-hadron jets. }
\label{fig:inclusive_jet_and_charged_jet}
\end{figure}

\begin{figure}[t]
\centering
\includegraphics[width=0.99\textwidth]{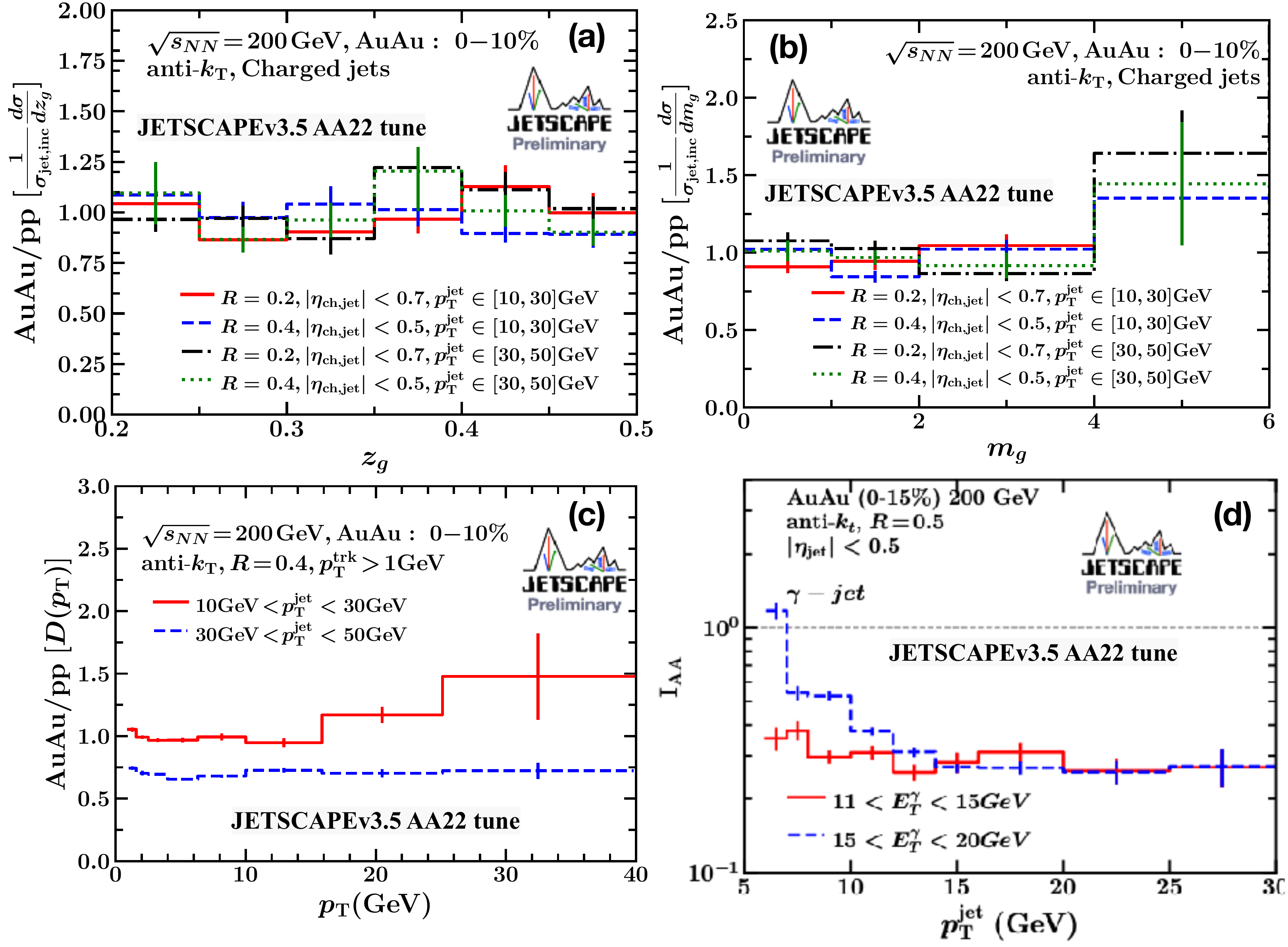}
\caption{
Nuclear modification of jets at most central ($0-10\%$) collisions at $\sqrt{s_{\mathrm{NN}}}=$200 GeV. The calculation is performed using JETSCAPEv3.5 AA22 tune described in Ref.~\cite{JETSCAPE:2022jer}. The parameters in the Soft Drop grooming algorithm are $z_{\mathrm{cut}}=0.2$ and $\beta=0$.
(a) The distribution of $z_{\mathrm{g}}=\mathrm{min}(p^{\mathrm{prong}}_{\mathrm{T}1},p^{\mathrm{prong}}_{\mathrm{T}2}) )/( p^{\mathrm{prong}}_{\mathrm{T}1} + p^{\mathrm{prong}}_{\mathrm{T}2}  )$ for charged-jets of cone size $R=$0.2 and 0.4.
(b) Groomed jet mass for charged jets of cone size $R=$0.2 and 0.4.
(c) $p_{\mathrm{T}}$ dependence of jet fragmentation function for inclusive jets of cone size $R=$0.4.
(d) Nuclear modification of photon-triggered inclusive jets of cone size $R=$0.5.
}
\label{fig:groomed_jets_fg_photon_jet}
\end{figure}

Fig.~\ref{fig:inclusive_jet_and_charged_jet} presents the nuclear modification factor of inclusive jets and charged jets for jet cone radius values $R$=0.4, 0.6, and 0.8.  The predictions show that the charged jets are more suppressed compared to inclusive jets at high jet $p_{\mathrm{T}}$ values. Also, in contrast to jets at LHC, RHIC jets do not show strong jet cone size dependence.
Next, predictions are presented for groomed charged-jet observables computed using the Soft Drop grooming algorithm with parameters $z_{\mathrm{cut}}=0.2 $ and $\beta$=0 in Fig.~\ref{fig:groomed_jets_fg_photon_jet}(a) and (b).
The prediction shows that the nuclear modification of $z_{\mathrm{g}}$ distribution is consistent with unity, indicating that there is no strong modification in the hardest splitting during in-medium jet evolution. Moreover, the trend looks similar to ALICE measurements at $\sqrt{s_{\mathrm{\mathrm{NN}}}}=$5.02 TeV. The predictions for jet mass [Fig.~\ref{fig:groomed_jets_fg_photon_jet}(b)] also shows no significant nuclear modification at low groomed jet mass $m_{g}<4$ GeV.

Fig.~\ref{fig:groomed_jets_fg_photon_jet}(c) presents predictions for fragmentation function [$D(p_{\mathrm{T}})$] as a function of hadron $p_{\mathrm{T}}$ for two different inclusive $p^{\mathrm{jet}}_{\mathrm{T}}$ ranges. The results show strong suppression for high $p_{\mathrm{T}}$ jets.
In the end, results are presented for nuclear modification ($I_{\mathrm{AA}}$) of photon-triggered inclusive jets [Fig.~\ref{fig:groomed_jets_fg_photon_jet}(d)]. The results indicate a rise in $I_{\mathrm{AA}}$ below $p^{\mathrm{jet}}_{\mathrm{T}}=12$ GeV, whereas it is strongly suppressed  for $p^{\mathrm{jet}}_{\mathrm{T}}>12$GeV and remains constant for higher  $p^{\mathrm{jet}}_{\mathrm{T}}$.

In addition to the specific predictions above, a large sPHENIX data-set could provide many additional opportunities to test the jet quenching model and probe the hardest in-medium splitting, path-length dependence of photon-triggered jets, and in-medium energy loss of quark and gluon jets.

\subsection{$R_{\rm AA}$ and $v_2$ as jet substructure observables with sPHENIX} 


Jets are multi-partonic systems that develop before interactions with the QGP set in and lead to energy loss and to substantial modifications of their substructure. 

\begin{figure*}[t]
    \centering
    \includegraphics[width=0.49\textwidth]{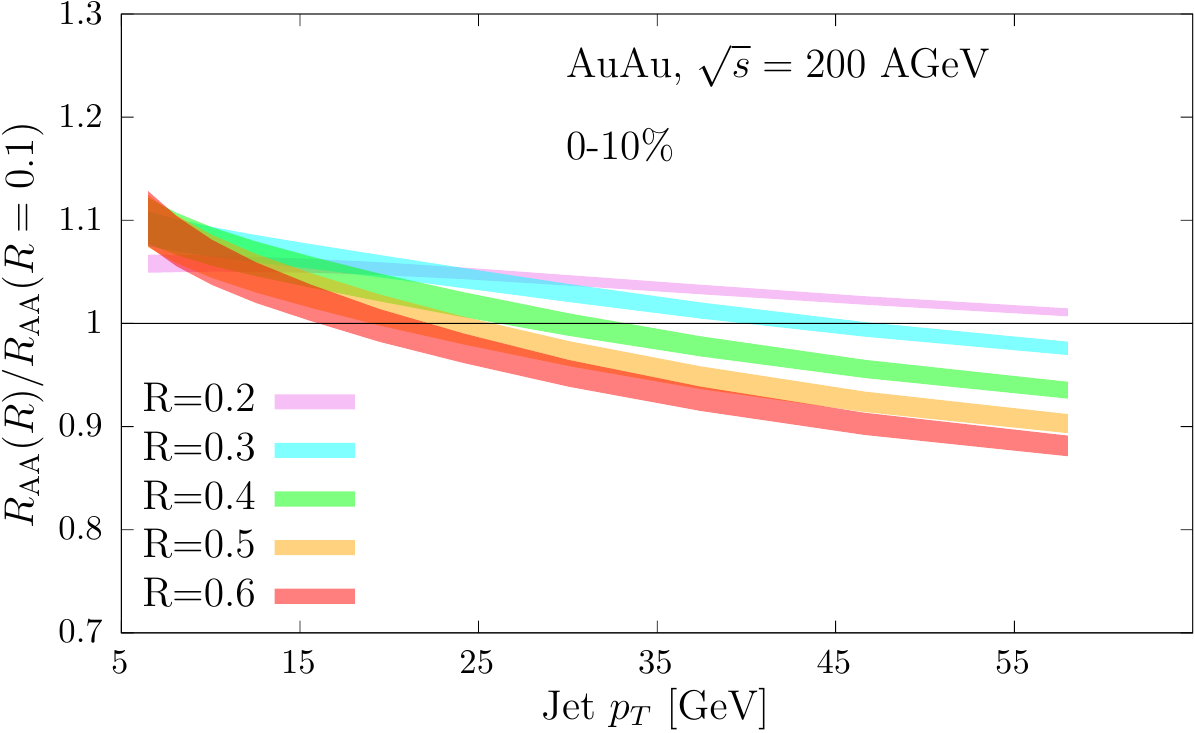}
    \includegraphics[width=0.49\textwidth]{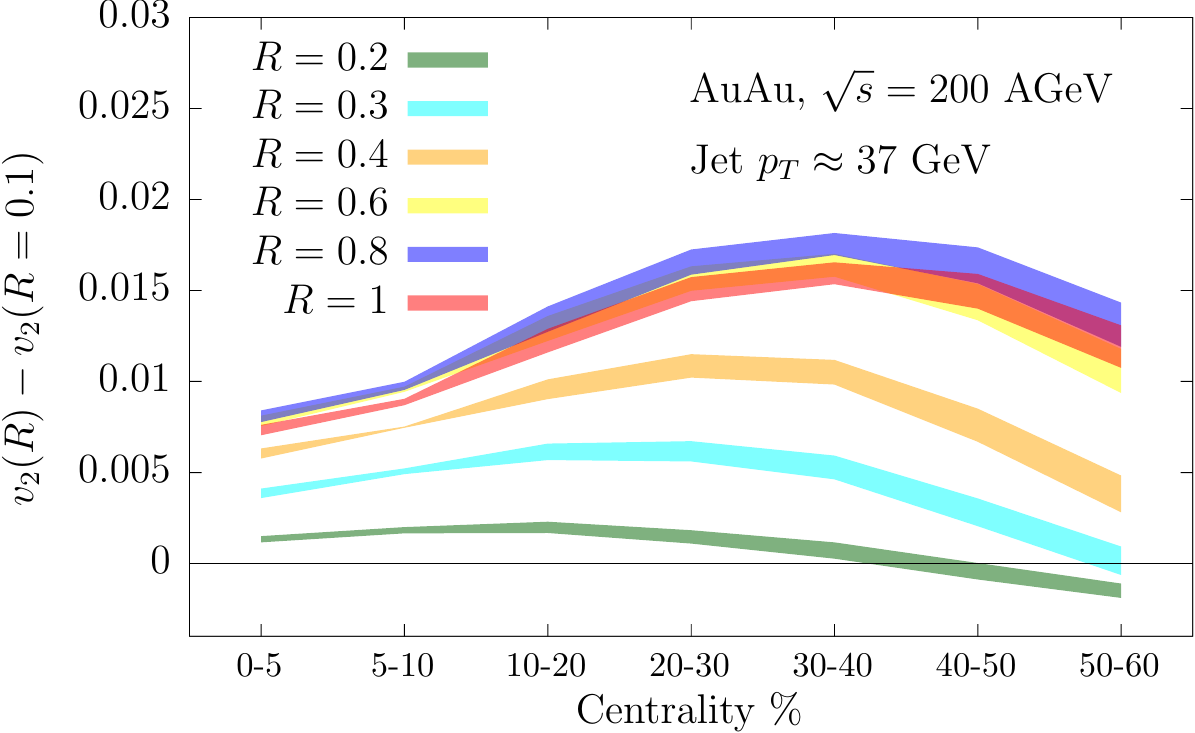}
    \caption{Left: Ratio of jet $R_{\rm AA}$ for different $R$ values to the $R_{\rm AA}$ with $R=0.1$. Right: Difference in $v_2$ between different $R$ and $R=0.1$, for a jet $p_T\approx 37$ GeV, as a function of centrality.}
    \label{fig:yacinedanikonrad}
\end{figure*}

Jet modification depends on the degree to which the medium can resolve the internal jet structure that is dictated by the physics of coherence governed by a critical angle $\theta_c$. To leading logarithmic accuracy, parton splittings that take place within $\theta<\theta_c$ will not be resolved by the medium. Therefore, in order to compute the total energy lost by a jet, one needs to consider not only where the radiated energy ends up, but also the resolved phase-space for a given jet $p_T$ and cone-size $R$ which would be affected by quenching effects. Using resummed quenching weights that incorporate the Improved Opacity Expansion (IOE) framework for medium-induced radiation~\cite{Barata:2020sav} and embedding the system into a realistic heavy-ion environment for AuAu collisions at $\sqrt{s}=200$ GeV, the $R$ dependence of jet suppression, $R_{\rm AA}(R)$, is computed in the left panel of Fig.~\ref{fig:yacinedanikonrad}.
The only two parameters of the model are $g_{\rm med}$, determining the (so far fixed) coupling between the energetic partons and the constituents of the QGP, and $R_{\rm rec}$, which estimates the extent to which thermalized energy is recovered as a function of $R$ within the jet hemisphere. In these preliminary results, $g_{\rm med}$ is varied within the range $g_{\rm med}\in \lbrace 2.3,2.4\rbrace$, and set $R_{\rm rec}=\pi/2$ (corresponding to a flat redistribution of the energy). The variation of $R_{\rm rec}$ has limited effect on jets up to $R\sim 0.6$.
Results are presented for jet suppression in terms of a ratio between $R>0.1$ and $R=0.1$.
Overall, there is a very mild $R$-dependence for the range of $R$ studied, similar to what was found at the LHC~\cite{Mehtar-Tani:2021fud}, with variations up to $\sim 10\%$.
At lower jet $p_T$, larger $R$ jets tend to be somewhat less suppressed than small-cone jets. This is due to the fact that the radiated energy is more likely to remain within the jet for larger cone size jets. However, by increasing jet $p_T$, the size of the jet phase-space increases more for larger $R$, and then the trend is reversed, with larger $R$ jets more quenched. 

Results are also presented for the jet azimuthal anisotropy $v_2(R)$ as a function of the difference between jets with $R>0.1$ and those with $R=0.1$, as a function of centrality, in the right panel of Fig.~\ref{fig:yacinedanikonrad}. Jet $p_T$ has been set to be approximately $p_T^{\rm jet}\approx 37$ GeV. The results show that as centrality is decreased, $v_2$ for moderate $R$ jets, such as $R=0.3$ and $R=0.4$, sequentially collapse towards the result for small $R=0.1$. The reason of this sequential grouping is the evolution of $\theta_c$ with centrality due to its strong dependence on the in-medium traversed length, $\theta_c\sim 1/\sqrt{\widehat{q}L^3}$. For those jets with $R>\theta_c$, traversing shorter lengths within the medium will make a larger difference than for those jets with $R<\theta_c$, since the size of the resolved phase-space over which quenching weights are resummed will be reduced. 
For this reason, $v_2(R)$ is quite sensitive to the typical value of $\theta_c$ at a given centrality. Further details on these results will be presented in a forthcoming publication. 

\subsection{CoLBT-hydro predictions for sPHENIX}


The CoLBT-hydro model \cite{Chen:2017zte,Chen:2020tbl,Zhao:2021vmu} is developed to simulate jet propagation and jet-induced medium response in high-energy heavy-ion collisions. It combines the microscopic linear Boltzmann transport (LBT) model \cite{He:2015pra} for the propagation of energetic jets and recoil partons with the event-by-event (3+1)D CCNU-LBNL viscous (CLVisc) hydrodynamic model \cite{Pang:2012he,Pang:2014ipa,Pang:2018zzo} for the evolution of the bulk medium and soft modes of the jet-induced medium response. LBT and CLVisc are coupled in real time through a source term from the energy-momentum lost to the medium by jet shower and recoil partons as well as the particle-holes or ``negative partons" from the back-reaction. The LBT model \cite{He:2015pra} is based on the Boltzmann equation for both jet shower and recoil partons with perturbative QCD (pQCD) leading-order elastic scattering and induced gluon radiation according to the high-twist approach \cite{Guo:2000nz,Wang:2001ifa,Zhang:2003yn,Zhang:2003wk}. The CLVisc~\cite{Pang:2014ipa,Pang:2012he,Pang:2018zzo}  viscous hydrodynamic model with the default  freeze-out temperature $T_f=137$ MeV,  specific shear viscosity $\eta/s=0.15$, the s95p parameterization of the lattice QCD EoS with a rapid crossover \cite{Huovinen:2009yb} and Trento \cite{Moreland:2014oya} initial conditions with a longitudinal envelope at an initial time $\tau_0=0.6$ fm/$c$ can reproduce experimental data on bulk hadron spectra and anisotropic flows at both RHIC and  LHC energies. The Trento model is also used to provide the transverse spatial distribution of jet production whose initial configurations are generated from PYTHIA8 \cite{Sjostrand:2007gs}. 
Partons from the initial jet showers, as well as MPI, propagate through the QGP and generate medium response according to the CoLBT-hydro model. The final hadron spectra include contributions from the hadronization of hard partons within a parton recombination model \cite{Han:2016uhh,Zhao:2020wcd} and jet-induced hydro response via Cooper-Frye freeze-out after subtracting the background from the same hydro event without the $\gamma$-jet.  More detailed descriptions of the LBT and CoLBT-hydro models are given in Refs.~\cite{He:2015pra,Cao:2016gvr,Cao:2017hhk,He:2018xjv,Luo:2018pto,Zhang:2018urd} and \cite{Chen:2017zte,Zhao:2021vmu,Yang:2021qtl}.

\begin{figure}[t]
\centering
\includegraphics[width=0.49\textwidth]{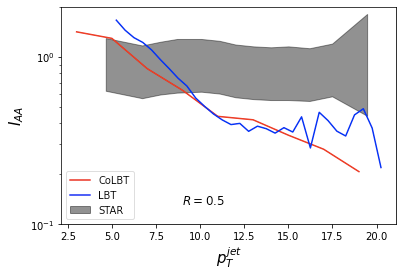}
\includegraphics[width=0.49\textwidth]{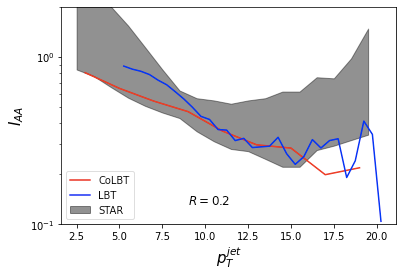}
\caption{Nuclear modification factor $I_{AA}$ as a function of $p_T^{\rm jet}$ for $\gamma$-jet with $p_T^\gamma$ = 15--20 GeV in central 0--10\% Au+Au collisions at $\sqrt{s}=200$ GeV from the CoLBT-hydro (blue) and LBT models, compared here to preliminary STAR data \cite{Sahoo:2020kwh}. }
\label{fig:iaa}
\end{figure}

\begin{figure}[t]
\centering
\includegraphics[width=0.49\textwidth]{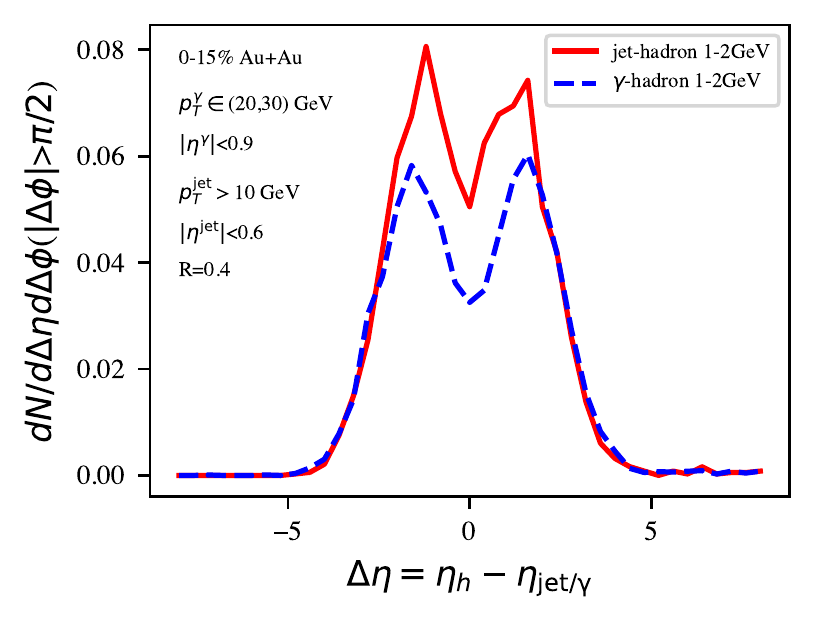}
\caption{$\gamma$-hadron (blue dashed) and jet-hadron (red solid) correlation  in 0--15\% Au+Au collisions at $\sqrt{s}=200$ GeV as a function of the rapidity difference $\eta_h-\eta_{\rm jet}$ in the $\gamma$ azimuthal direction.}
\label{fig:diffusion}
\end{figure}

Shown in Fig.~\ref{fig:iaa} are the nuclear modification factors  of the $\gamma$-jet process as a function of $p_T^{\rm jet}$ in 0--15\% central Au+Au collisions at $\sqrt{s}$ = 200 GeV from CoLBT-hydro  (red) and LBT (blue) model simulations as compared to the STAR preliminary data \cite{Sahoo:2020kwh} for two different jet cone size values of $R=0.5$ and 0.2. Both model results show a weak jet cone size dependence. Shown in Fig.~\ref{fig:diffusion} are $\gamma$-hadron (dashed blue) and jet-hadron correlations (red solid) for soft charged hadrons ($p_T$ = 1--2 GeV/$c$) as a function of the rapidity difference $\eta_h-\eta_{\rm jet}$ in the $\gamma$ azimuthal directions. The dip on top of the MPI Gaussian peak is caused by jet-induced diffusion wake. The dip becomes deeper for smaller $\gamma$-jet asymmetry $p_T^{\rm jet}/p_T^\gamma$ or bigger jet energy loss. These phenomena can be explored with sPHENIX data.

\subsection{Disentangling jet modification in sPHENIX}


The selection of jets in heavy-ion collisions based on their $p_T$ after jet quenching is known to bias towards jets that lost little energy in the quark-gluon plasma. The work in Ref.~\cite{Brewer:2021hmh} studied and quantified the impact of this selection bias on jet
substructure observables so as to isolate effects caused by the modification of the substructure of jets by quenching. This study was performed in a simplified Monte Carlo setup, in which it was possible to identify the same jet before and after quenching.
The work then showed explicitly that jets selected based on their quenched (i.e. observable) $p_T$ have substantially smaller fractional energy loss than those selected based on the $p_T$ that they would have had in the absence of any quenching. This selection bias has a large impact on jet structure and substructure observables. As an example, one can consider
the angular separation $\Delta R$ of the hardest splitting in each jet, and found that the $\Delta R$ distribution of the (biased) sample of jets selected based upon their quenched $p_T$ is almost unmodified by quenching. In contrast, quenching causes dramatic modifications to the $\Delta R$ distribution of a sample of jets selected based upon their unquenched $p_T$, with a significant enhancement at larger $\Delta R$ coming from the soft particles originating from the wake of the jet in the quark-gluon plasma. The jets which contribute to this enhancement are those which have lost the most energy and which were, therefore, left out of the sample selected after quenching. In the second part of Ref.~\cite{Brewer:2021hmh}, a more realistic study showed that the same qualitative effects could all be observed in events where a jet is produced in association with a $Z$ boson at the LHC. Selecting jets in such events based on either the jet $p_T$ or the $Z$-boson $p_T$ provides an experimentally accessible way to quantify the effects of selection biases in jet observables and separate them from the modification of jet substructure caused by quenching. 
In this section, results are presented from repeating the same study for the sPHENIX experiment at RHIC using events in which 
a jet is produced in association with a direct photon.

\begin{figure}[t]
\centering
\includegraphics[width=0.45\textwidth]{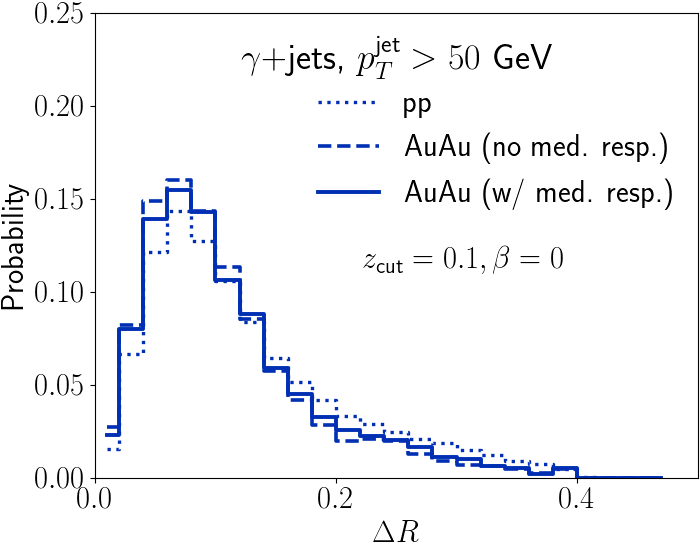}\\
\includegraphics[width=0.45\textwidth]{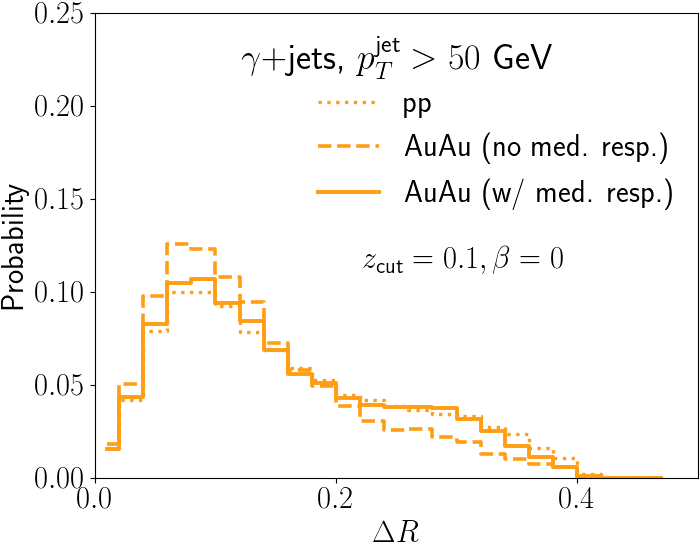}
\includegraphics[width=0.45\textwidth]{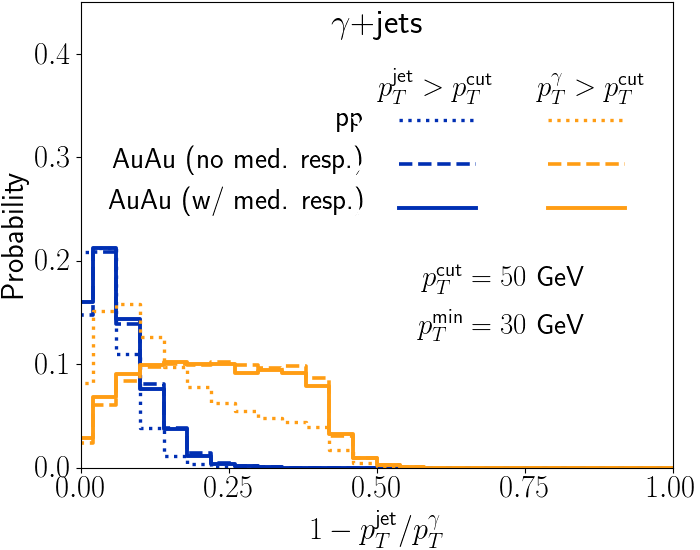}
\caption{
		Top: Fractional $p_T$ asymmetry between a photon and its recoiling jet, for jets with jet $p_T$ above $50$~GeV (blue) and photon $p_T$ above $50$~GeV (orange) in vacuum (dotted), and in heavy ion collisions with hadrons originating from medium response artificially excluded from (dashed) or included in (solid) the reconstructed jets. Bottom left and right panels show the distribution of $\Delta R$ for the same samples of jets.}
\label{fig:fig6}
\end{figure}

The results below are based on simulations of jets produced in association with a direct photon with pseudorapidity $|\eta^\gamma|\le0.9$ in Au+Au collisions at $\sqrt{s} = 200$~GeV in the Hybrid Model of jet quenching~\cite{Casalderrey-Solana:2014bpa}.
It is experimentally challenging to access direct photons at low $p_T^\gamma$, due to backgrounds from photons that are not produced in the hard process. 
This study assumes that sPHENIX will be able to isolate a clean sample of jets produced in association with direct photons down to $p_T^\gamma = 30$~GeV, and will have sufficient statistics to access hard processes with $p_T \ge 50$~GeV. 
As in Ref.~\cite{Brewer:2021hmh} jets are selected in two distinct ways, which yield results that are qualitatively different from one another in interesting ways. Fig.~\ref{fig:fig6} shows the $p_T$ asymmetry and $\Delta R$ distributions of jets in this sample selected in two ways. The blue histogram shows jets with $p_T^\mathrm{jet} \ge 50$~GeV recoiling against a photon satisfying $p_T^\gamma \ge 30$~GeV, while the orange histogram shows jets with $p_T^\mathrm{jet} \ge 30$~GeV recoiling against a photon with $p_T^\gamma \ge 50$~GeV. Fig.~\ref{fig:fig6}(a) illustrates that jets selected using the first method lose a much smaller fraction of their energy.
This selection bias is also born out in the substructure of jets selected in each way. The $\Delta R$ distribution of jets with $p_T^\mathrm{jet} \ge 50$~GeV (Fig.~\ref{fig:fig6}(b)) is much narrower than for those produced in association with a photon with $p_T^\gamma \ge 50$~GeV (Fig.~\ref{fig:fig6}(c)), due to the selection bias in the former case against jets that lost more energy (typically, those with larger $\Delta R$). As in Ref.~\cite{Brewer:2021hmh}, there is a (small) enhancement at large $\Delta R$ in Fig.~\ref{fig:fig6} coming from the response of the medium to jets. However, this effect is more muted than at the LHC, presumably due to the smaller separation between the minimum $p_T$ of direct photons and the statistical reach of the hard process at RHIC. This effect could be enhanced in the scenario that sPHENIX is able to access photons of lower $p_T^\gamma$ or jet substructure of jets with higher $p_T$. This possibility should be explored more systematically in future work and as the specific kinematic capabilities in the recorded data become more clear.

\subsection{Comprehensive SCET$_\mathrm{G}$ predictions for sPHENIX}


Predictions for the sPHENIX program are presented, including the modification of light and heavy-flavor hadrons and jets~\cite{Kang:2014xsa,Ke:2022gkq,Vitev:2009rd,Huang:2013vaa}, photon-tagged jets and dijets~\cite{Dai:2012am,Kang:2018wrs}, and jet substructure~\cite{Vitev:2009rd,Li:2017wwc}. 

\begin{figure}[t]
\includegraphics[width=0.45\textwidth]{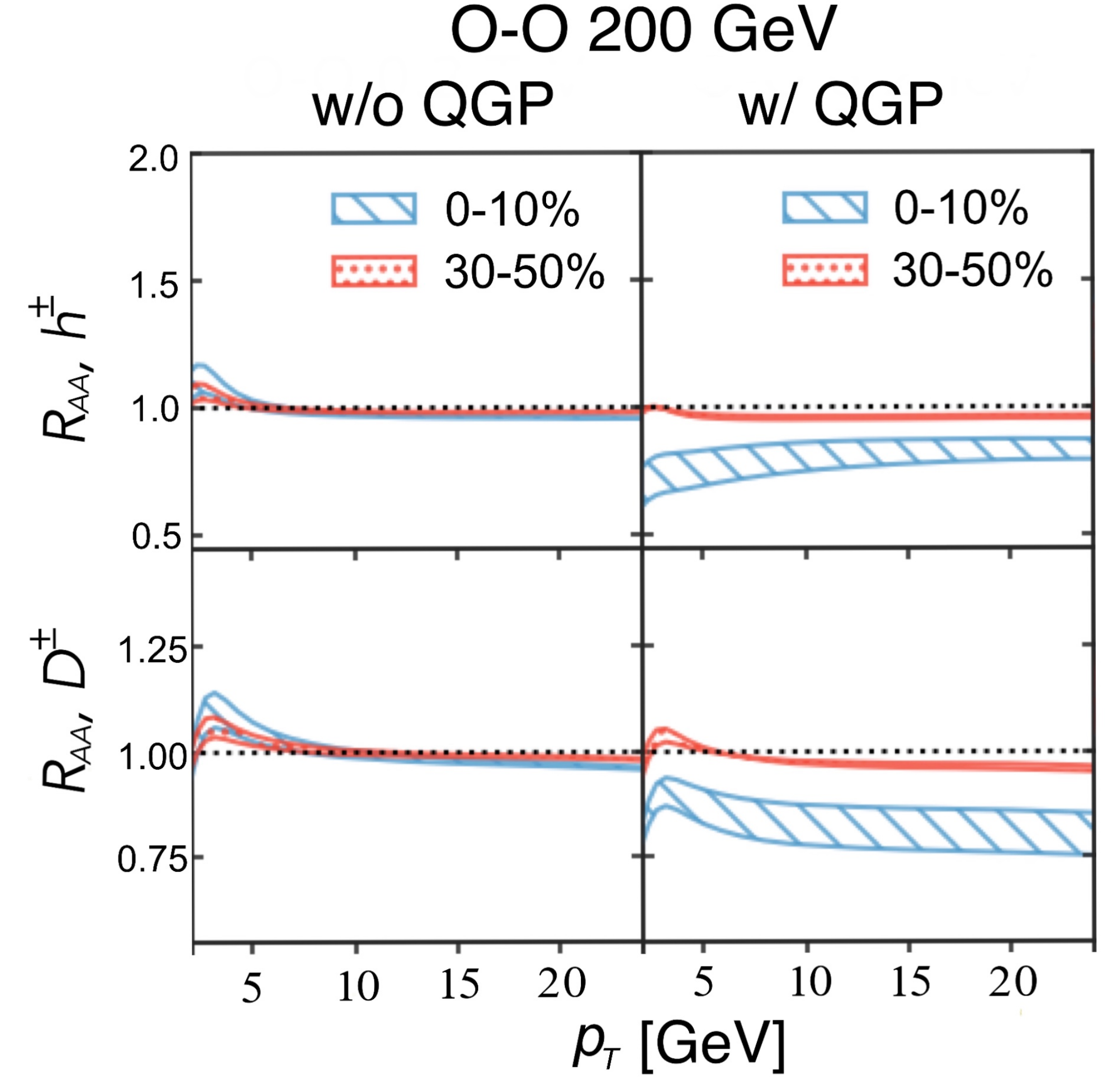}
\includegraphics[width=0.53\textwidth]{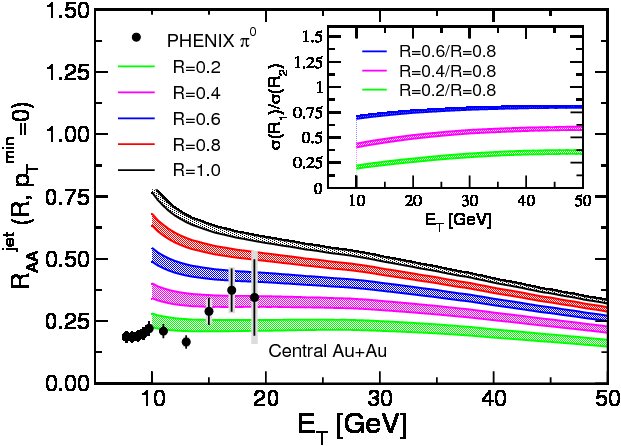}
\caption{Left: Nuclear modification of light and heavy flavor without (left column)
 and with (right column) QGP effects  in 0-10\%  and 30-50\% O+O  collisions at $\sqrt{s_{NN}}=200$~GeV. Right: Transverse energy dependent nuclear modification factor $R_{AA}^{\rm jet}$
for different cone radii $R$ in 0-10\% Au+Au  collisions at $\sqrt{s_{NN}}=200$~GeV. Inserts show ratios of jet cross 
sections for different $R$ in nuclear reactions versus  $E_T$. \label{Vitev1}}
\end{figure}

The essential  prediction for hard probes in heavy ion collisions is the suppression of the cross sections of  light and heavy hadrons and jets. The suppression of hadrons can be addressed in both the traditional energy loss approach and using modern QCD evolution techniques, based on the ability to calculate full in-medium splitting functions in dense nuclear matter~\cite{Kang:2014xsa}. While phenomenologically the results are similar, the latter  approach constitutes a more rigorous and improvable treatment of the problem. The suppression of light hadrons in central Au+Au collisions is predicted to be approximately  a factor of four to five. At high transverse momenta $p_\mathrm{T} > 20$~GeV cold nuclear matter (CNM) effects play an important role. At small transverse momenta, CNM effects are small and can only be observed in $p(d)$+Au collisions where they lead to Cronin-like enhancement, which is more pronounced at backward rapidity. An important question that is still unresolved is regarding CNM vs QGP effects in small systems. Recent calculations~\cite{Ke:2022gkq} predict that the difference is most pronounced in small but symmetric systems, such as O+O in the left of Fig.~\ref{Vitev1}. This can be seen for both light ($h^\pm$) and heavy ($D^\pm$) flavor particles. Reconstructed inclusive jets  show similar suppression to light hadrons in the small radius  $R=0.2$ limit, see the right panel of Fig.~\ref{Vitev1}. A clear reduction of suppression is predicted as $R$ grows.  For heavy-flavor-tagged jets, such as $b$-jets, sPHENIX is expected to have sensitivity to the dead cone effect~\cite{Huang:2013vaa}, and the reduction of the suppression due to the heavy quark mass is found to play a role for $p_T < 25$~GeV. Furthermore, while the presence of gluon to heavy-flavor fragmentation processes was found to play an important role at the LHC, leading to similar light and heavy-flavor-tagged jet suppression, the smaller expected gluon contribution at RHIC energies is instead predicted to result in slightly smaller suppression for $b$-jets than for inclusive jets. 
           
\begin{figure}[t]
\includegraphics[width=0.51\textwidth]{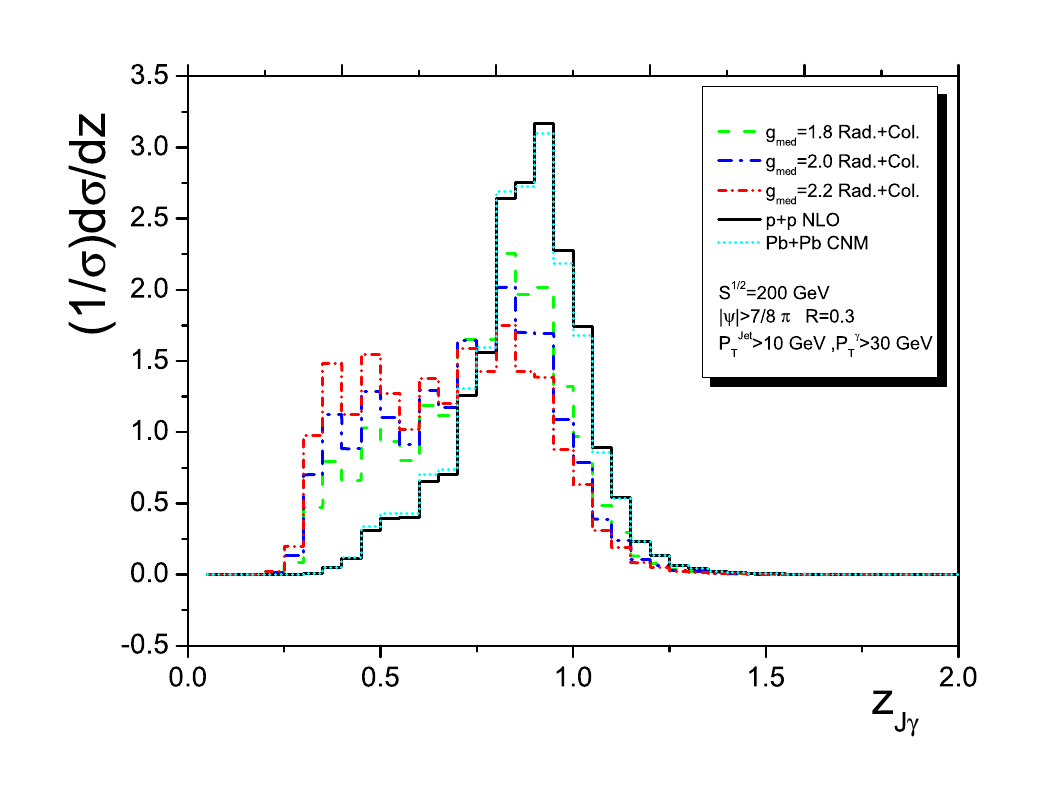}
\includegraphics[width=0.48\textwidth]{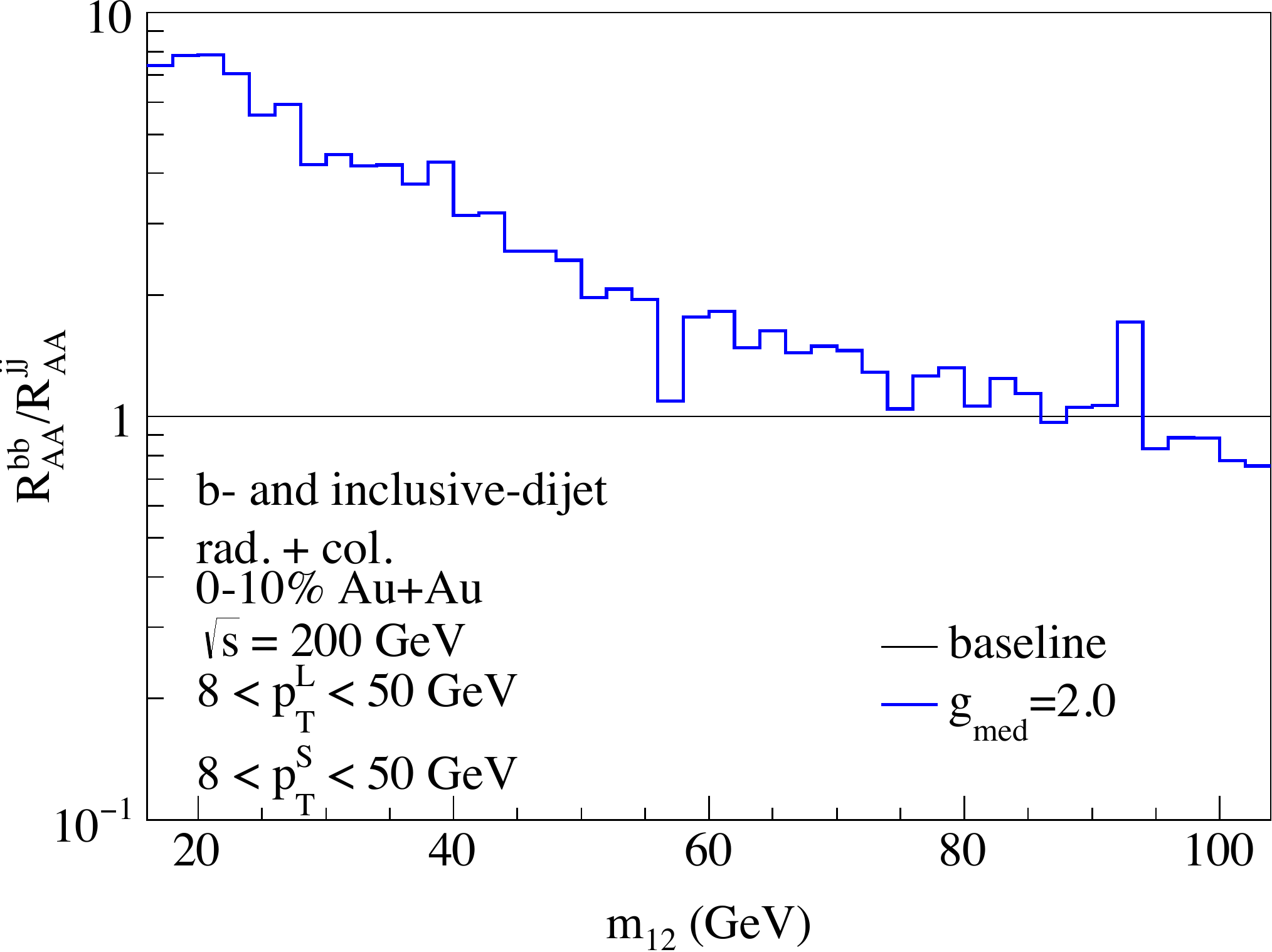}
\caption{Left: The isolated photon-tagged jet asymmetry distribution for different
coupling  strengths between the jet and the medium for Au+Au collisions at $\sqrt{s_{NN}} = 200$~GeV. Right: Ratios of nuclear modification factors for $b$-tagged ($R_{AA}^{bb}$) v.s inclusive ($R_{AA}^{jj}$) dijet production sPHENIX (right) are plotted as a function of dijet invariant mass $m_{12}$. \label{Vitev2}}
\end{figure}

Photon-tagged jets and jet correlations provide more differential probes of in-medium dynamics. The advantage of photon tagging is that it can provide a reference  relative to which the recoil  jet energy loss can be studied. One important prediction is that because of the non-monotonic form of the photon-jet cross section, the modification factor (sometimes called $I_{AA}$) has a non-trivial $p_T$  dependence where the suppression decreases near the trigger photon momentum and can even turn into enhancement~\cite{Dai:2012am}.  In addition, the calculations predict a significant modification of the photon-tagged jet imbalance $z_{J\gamma}$ distribution, shown in the left panel of Fig.~\ref{Vitev2}. Since the reference $p$+$p$ distribution is narrower than at the LHC, the broadening is more pronounced and the  shift in the mean momentum imbalance $\Delta \langle z_{J\gamma} \rangle =   \langle z_{J\gamma} \rangle_{pp} -  \langle z_{J\gamma} \rangle_{AA}$ is very sensitive to the jet-medium coupling. One important task at RHIC will be to enhance the effects of jet quenching. The dijet mass (the mass of the dijet system) is a newly proposed observable that can achieve this goal~\cite{Kang:2018wrs}. It was found that the suppression in the mass $m_{12}$ distribution of light dijets  can differ from unity by an order of magnitude. At the same time this observable enhances the sensitivity to the heavy quark mass effect, such that the transverse momentum dependence of  $b$-dijet suppression is very different than the one for light dijets. This can be seen in the ratio of the nuclear modifications shown in the right panel of Fig.~\ref{Vitev2}   where the differences due to the $b$-quark  mass can be significant at low $m_{12}$~\cite{Kang:2018wrs}. 

\begin{figure}[t]
\includegraphics[width=0.49\textwidth]{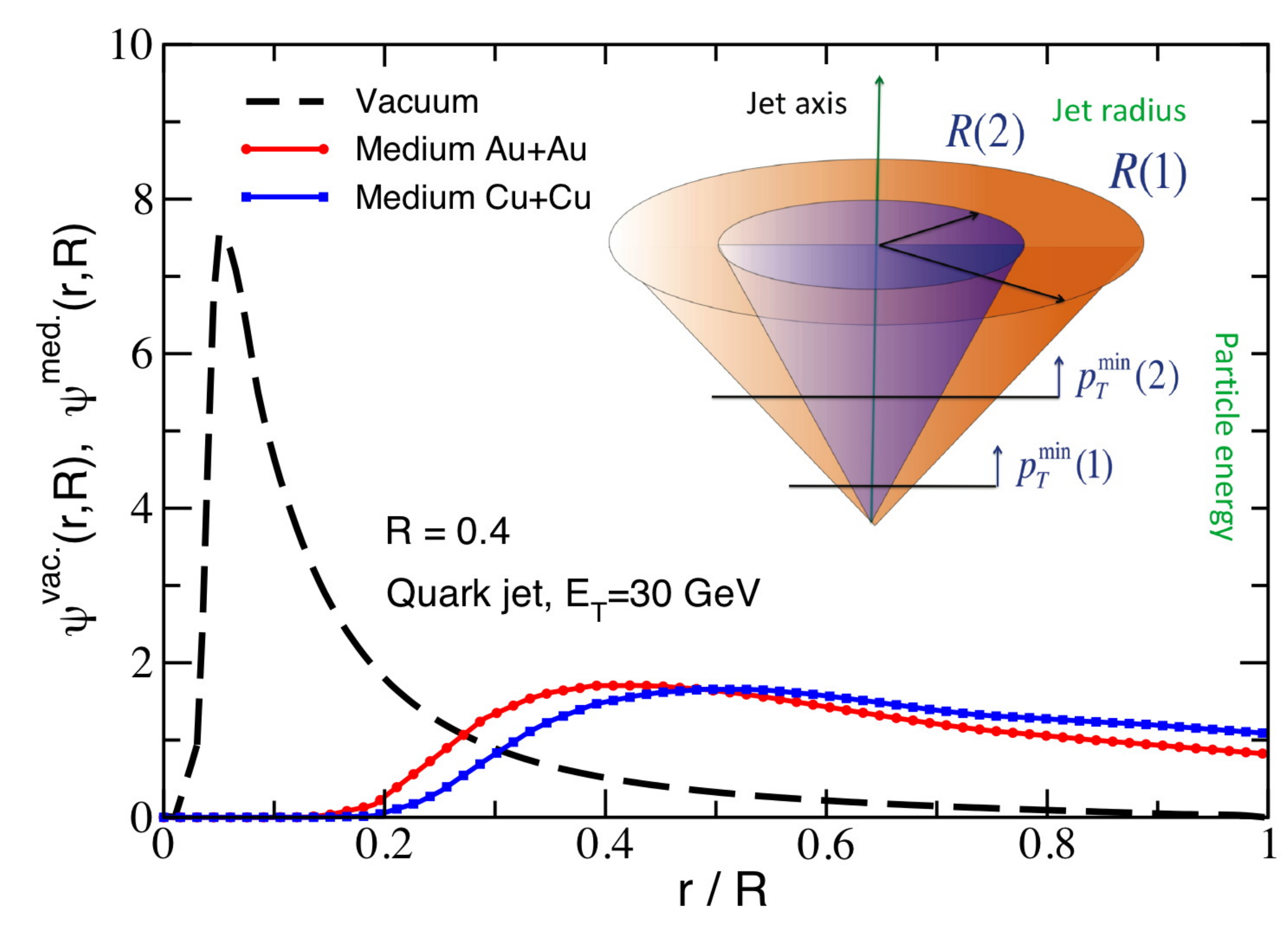}
\includegraphics[width=0.49\textwidth]{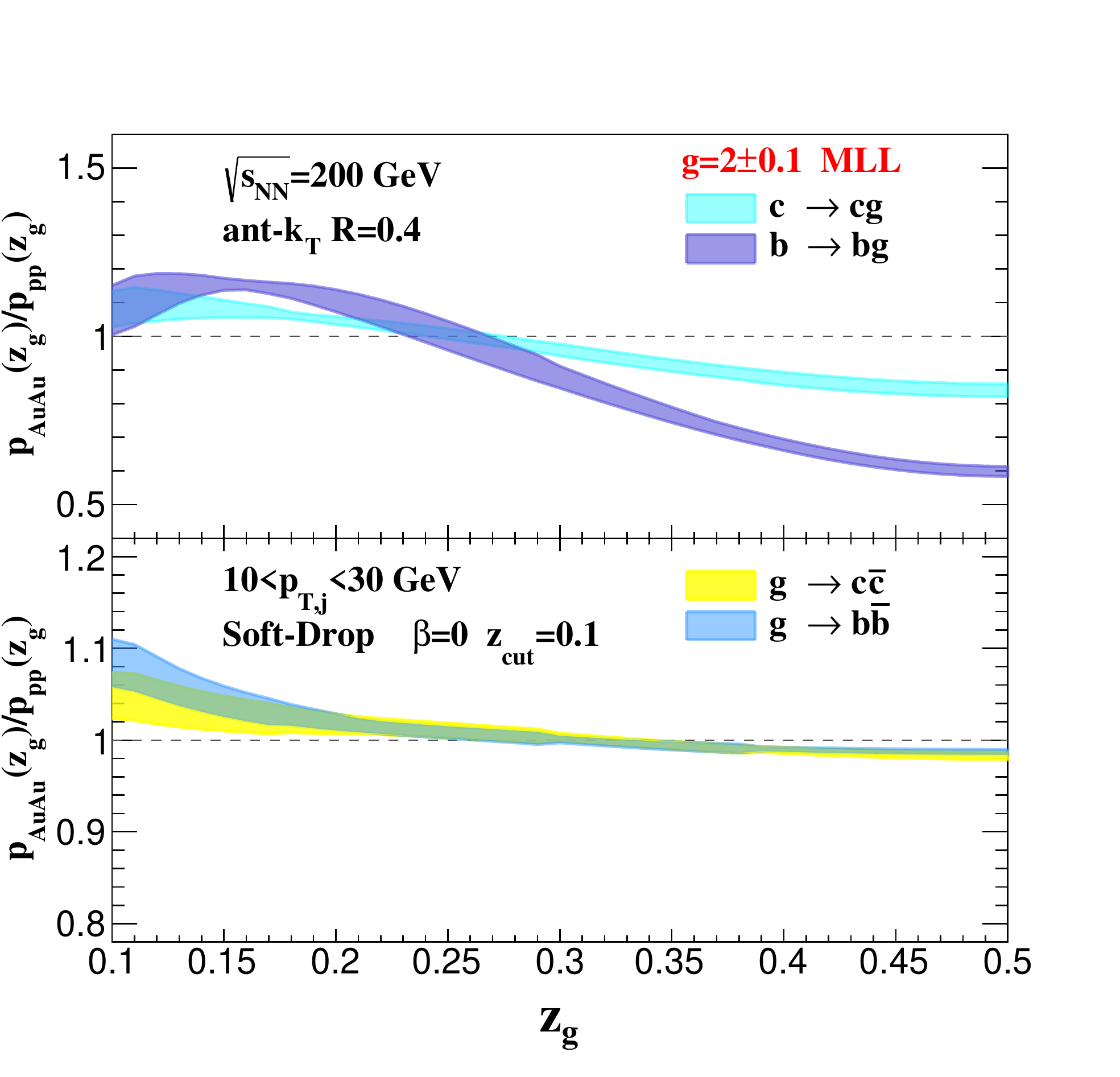}
\caption{Left: The differential jet shape in vacuum $\psi^{\rm vac.}(r,R)$ is contrasted to the  medium-induced 
contribution  $\psi^{\rm med.}(r,R)$  by a  $E_T = 30$~GeV quark in Au+Au and Cu+Cu collisions 
at $\sqrt{s_{NN}}=200$~GeV. The insert illustrates a method for studying the
characteristics of these parton showers. Right: The modifications of the splitting functions for heavy-flavor-tagged jet is shown for $\sqrt{s_{\rm NN}}=200$~GeV Au+Au collisions. An important feature is the strong quenching effects for prompt
    $b$-jets contrasted by the lack of QGP-induced modification for the $g\to Q\bar{Q}$ splitting.\label{Vitev3}}
\end{figure}

Jet substructure is a direct way to probe the properties of in-medium parton showers.  These studies were pioneered at the LHC and it is expected that sPHENIX will make valuable contributions to this physics at RHIC. In the left  panel of Fig.~\ref{Vitev3} the vacuum and medium-induced differential jet shapes for quark jets are shown~\cite{Dai:2012am}.  In comparison to the LHC, where gluons dominate, RHIC will be dominated by quark jets, which are narrower than the gluon jets. Thus, the enhancement of the jet shape toward the periphery of the jet $r/R \rightarrow 1$ is expected to be larger than at the LHC even if the narrow ``core'' region remains unmodified. 
Another  important substructure observable is the distribution of the soft dropped momentum-sharing fraction $z_g$,  and it is particularly  interesting for heavy-flavor jets \cite{Li:2017wwc}. For inclusive jets, the modification is much smaller than at the LHC. Going to lower jet transverse momenta  leads to a unique dependence of the jet momentum sharing distribution modification in heavy ion collisions - an  inversion of the mass hierarchy of jet quenching effects, which can be tested by sPHENIX.  This is shown in the right panel of Fig.~\ref{Vitev3} , where  ratios of the momentum sharing distribution for heavy-flavor tagged jets in  Au+Au to $p$+$p$ collisions  at $\sqrt{s_{\rm NN}}=200$ GeV are presented for  $10<p_{T, j}<30$~GeV  jets.  These corroborate the analytic expectations, showing that the magnitude of the effects is not only large, but particularly so for $b$-quark jets.   

\subsection{Energy correlations in jets with sPHENIX}


This calculation is based on a novel approach to jet substructure in heavy-ion collisions formulated in terms of energy correlation functions \cite{Andres:2022ovj}. It can be demonstrated that the scales of the QGP can be isolated as distinct features in the correlator spectra. For this purpose, the authors analyse the  case of the 2-point correlator $\langle \cE(\vec n_1) \cE(\vec n_2) \rangle$, which introduces a single scale-sensitive angular parameter, $\cos \theta =\vec n_1 \cdot \vec n_2 $. The $n$-th  weighted normalised 2-point correlator can be computed from the inclusive cross-section, $\sigma_{ij}$, to produce two hadrons ($i,j$) as

\begin{equation}
    \frac{\td \Sigma^{(n)}}{\td \theta} = \frac{1}{\sigma} \int \td \vec n_{1,2} \sum_{ij} \int \frac{\td \sigma_{ij}}{\td \vec n_i \td \vec n_j } \frac{E^{n}_{i} E^{n}_{j}}{Q^{2n}} \delta^{(2)}(\vec n_i - \vec n_{1})\delta^{(2)}(\vec n_j - \vec n_{2}) \delta(\vec n_2 \cdot \vec n_{1} - \cos \theta), 
\end{equation}
where $E_{i}$ is the lab-frame energy of hadron $i$, $\vec n_i$ is a two-component vector specifying its direction, and $Q$ is an appropriate hard scale. The 2-point correlator (EEC) is computed for a quark jet which propagates through a static medium of finite length $L$ and jet-quenching parameter $\hat q$ within the particular implementation of BDMPS-Z formalism~\cite{Baier:1996kr,Baier:1996sk,Zakharov:1996fv,Zakharov:1997uu} for semi-hard splittings given in \cite{Dominguez:2019ges,Isaksen:2020npj}. Vacuum collinear radiation is resummed using the celestial operator product expansion \cite{Hofman:2008ar,Konishi:1979cb,Hofman:2008ar,Dixon:2019uzg,Kologlu:2019mfz}. The result is that $\td \Sigma^{(n)}$ can be written in the factorised form \cite{Andres:2022ovj}:

\begin{equation}
    \frac{\td \Sigma^{(n)}}{\td \theta} = \frac{1}{\sigma_{qg}}\int \td z   \left(g^{(n)} + F_{\rm med} \right) \frac{\td \sigma^{\rm vac}_{qg}}{\td \theta \td z }  \label{eq:masterequation} z^n (1-z)^n \left(1  + \mathcal{O}\left(\As \ln \theta_{\TT{L}}^{-1}, ~\frac{\mu_{\rm s}}{z E} \right) \right) + \mathcal{O}\left(\frac{\mu_{\rm s} }{E} \right),
\end{equation}
where $g^{(1)}(\theta,\As) = \theta^{\gamma(3)} + \mathcal{O}(\theta)$ at fixed coupling given $\td\sigma^{\rm vac}_{qg}$ at $\mathcal{O}(\alpha_s)$. Here $\gamma(3)$ is the twist-2 spin-3 QCD anomalous dimension. $F_{\rm med}(z,\theta)$ is the medium-induced modification given in \cite{Isaksen:2020npj}, and $\td \sigma^{\rm vac}_{qg}$ is the vacuum splitting cross-section. $\mu_{\mathrm{s}}$ is the low scale of radiation over which $F_{\rm med}$ is inclusive. $Q$ is fixed as $Q=E$, the initial jet energy, and let $z = E_{g}/E$.

\begin{figure*}
\centering
\includegraphics[width=0.45\textwidth]{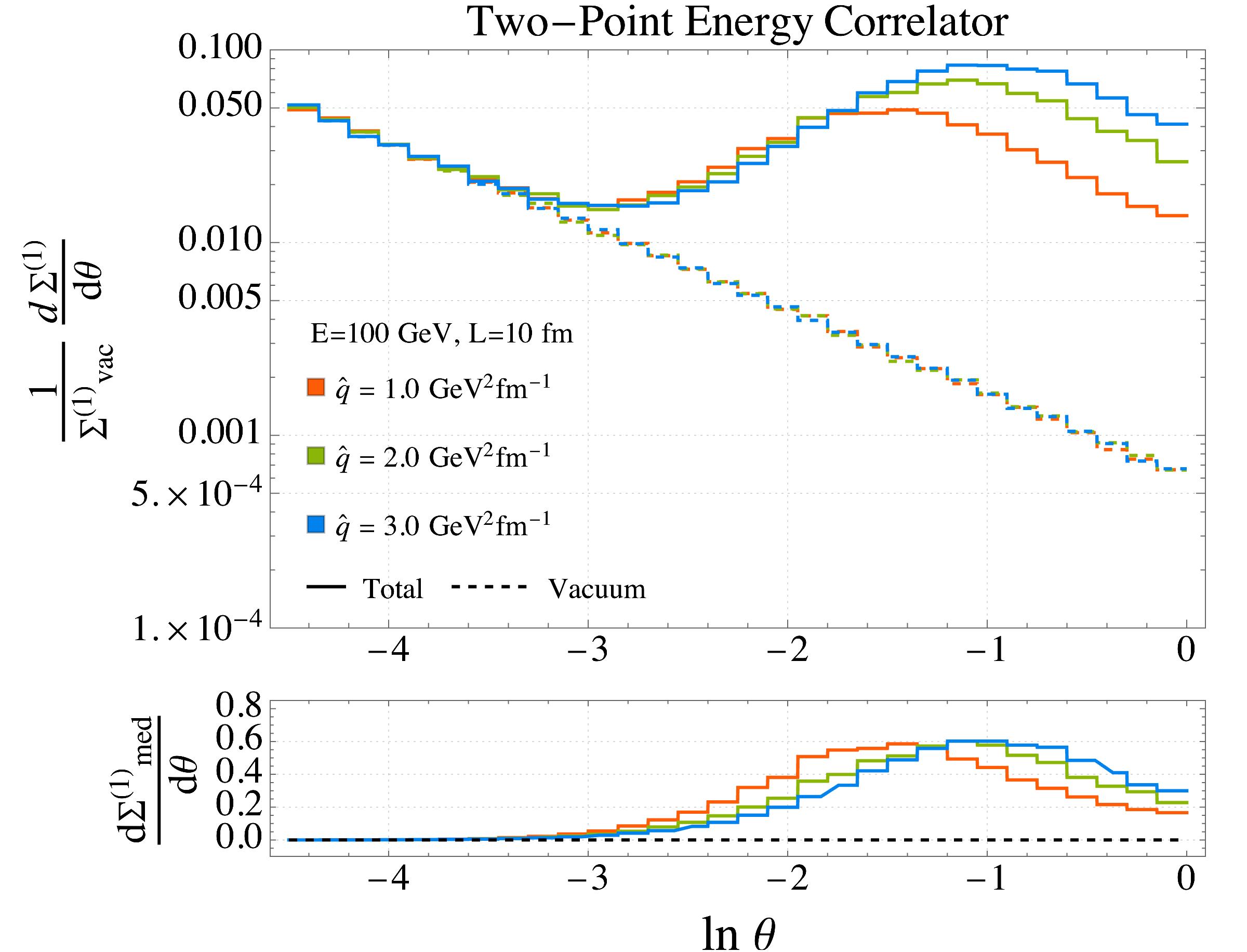}
\includegraphics[width=0.45\textwidth]{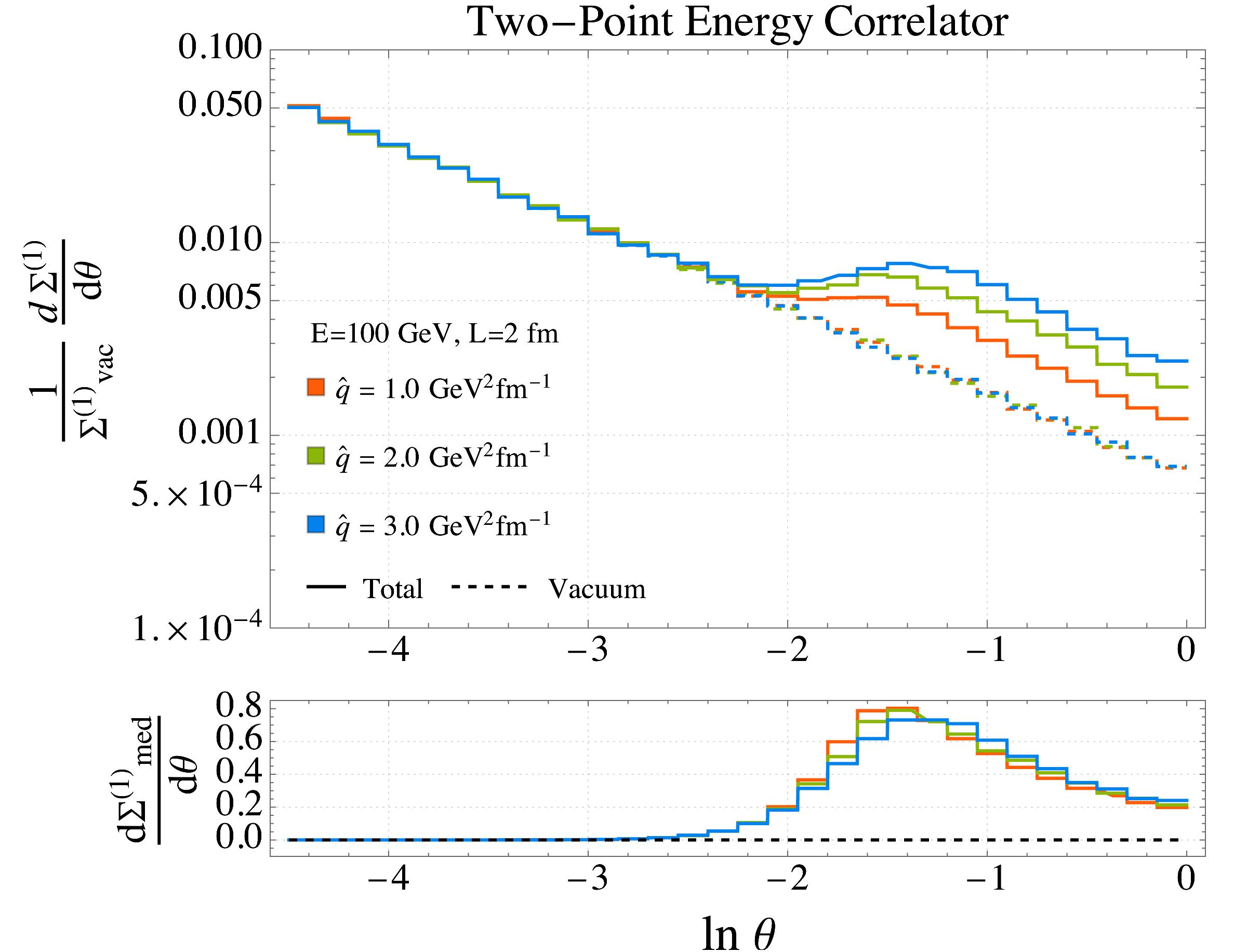}
\caption{The EEC evaluated using Eq.~\ref{eq:masterequation} for the DC (left panel) and PC regimes (right panel). The bottom panels show the volume normalised medium contribution to the distribution, defined as $\td \Sigma^{(n)}_{\TT{med}} = (\td \Sigma^{(n)} - \td \Sigma^{(n)}(\hat{q}=0))/\Sigma^{(n)}_{\TT{med}}$, so the shape can be more easily compared.}
  \label{fig:EEC100GeV}
\end{figure*}

\begin{figure}
\centering
\includegraphics[width=0.60\textwidth]{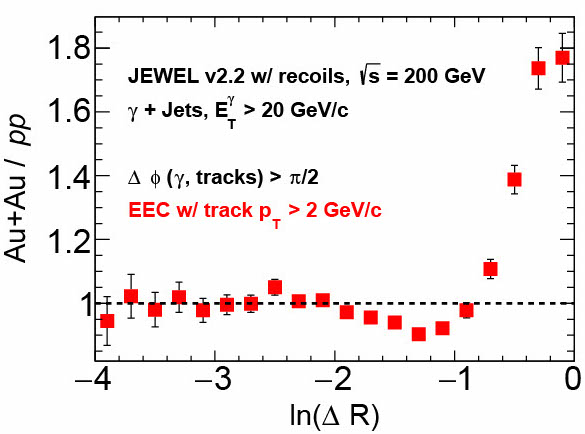}
  \caption{The EEC in \textsc{JEWEL}  with recoils for $\gamma$+jet Au$+$Au  events at $\sqrt{s_{\rm NN}} = 200$\,GeV and with a $350$ MeV medium temperature. A similar enhancement is found as in Fig.~\ref{fig:EEC100GeV}, however at larger $\Delta R$. This is consistent with the reduction in the jet energy of the sample from $100$~GeV in Fig.~\ref{fig:EEC100GeV} to $E^{\gamma}_{\rm T}>20$~GeV.
  While the specific predictions are shown here for $E=100$~GeV, the qualitative features are also expected for $E>30$~GeV partons at RHIC.
  }
  \label{fig:jewel}
\end{figure}

Fig.~\ref{fig:EEC100GeV} presents a numerical evaluation of Eq.~\ref{eq:masterequation}, where the parameters ($E,L,\hat{q}$) have been chosen such that the left panel  corresponds to a limit where one expects the quark-gluon pair to propagate decoherently (DC) through the medium. At the time of the workshop, the full calculations were performed only for $E = 100$~GeV partons. We have found qualitatively similar results with jet energies down to $E = 30$~GeV which are appropriate for RHIC kinematics, although these will need a full, quantitative treatment to understand the specific physics effect. 

The right panel corresponds to the regime where one expects the quark-gluon pair to propagate partially coherently through the medium (PC limit). A qualitatively different shape in the spectrum is readily observed between the two limits.  Ref.~\cite{Andres:2022ovj} demonstrates with a simple procedure that this difference can be used to extract the energy scale at which the onset of coherence occurs, i.e. the ``resolution scale'' of the QGP. This preliminary calculation highlights the exceptional potential of correlators to identify the presence of particular QGP dynamics at a given scale. In complement to this analysis, the 2-point correlator was also evaluated using the Monte Carlo parton shower JEWEL with recoils \cite{Zapp:2008gi,Zapp:2012ak,Zapp:2013vla,KunnawalkamElayavalli:2017hxo} for quark jets recoiling off photons \cite{KunnawalkamElayavalli:2016ttl}, as shown in Fig.~\ref{fig:jewel}. An enhancement at wide angles similar to that found in the (semi-)analytical analysis is clearly seen. The EEC was also found to be robust to a $2$\,GeV cut on the track $p_{T}$, as that typically used experimentally to suppress backgrounds. These measurements of the EEC can be performed using future sPHENIX data.

\subsection{Bayesian inference with JETSCAPE using sPHENIX data}


Bayesian inference provides a mechanism for systematic and agnostic physics interpretation of the wealth of information contained in jet quenching measurements at RHIC and the LHC. 

The JETSCAPE collaboration previously carried out a proof-of-principle Bayesian inference analysis using the inclusive charged hadron $\Raa{}$ at \(\sqrts{} =\) 200 GeV at RHIC and \(\sqrts{} =\) 2.76
and 5.02 TeV at the LHC~\cite{JETSCAPE:2021ehl}.
This analysis provided new constraints on the dependence of the jet transport coefficient $\qhat{}$ on medium temperature and parton momentum for several physics models, thereby demonstrating the viability of Bayesian inference to study jet quenching.

The next step is to expand the analysis to utilize multiple observables as a multi-messenger description of jet quenching, which encodes the structure and dynamics of the QGP into modifications of jet observables.
However, in this study only one new observable is added in order to isolate the impact of the additional information.
To this end, a new Bayesian inference was carried out with inclusive jet and charged hadron $\Raa{}$, including all available experimental data for fully corrected inclusive jet distributions.
This analysis utilizes a $\qhat{}$ parametrization with six parameters, and reduces the number of interactions at high virtuality due to coherence effects.
Details of the model are reported in Sec. 
\ref{JETSCAPE_kumar} above, and are further described in \cite{JETSCAPE:2022jer}.

\begin{figure}[t]
    \centering
    \includegraphics[width=0.8\textwidth]{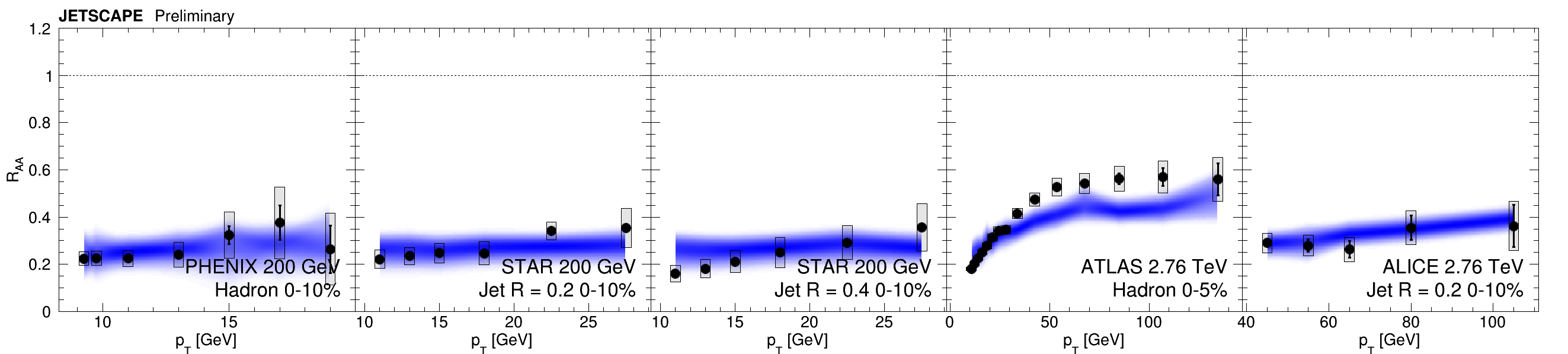}
    \includegraphics[width=0.8\textwidth]{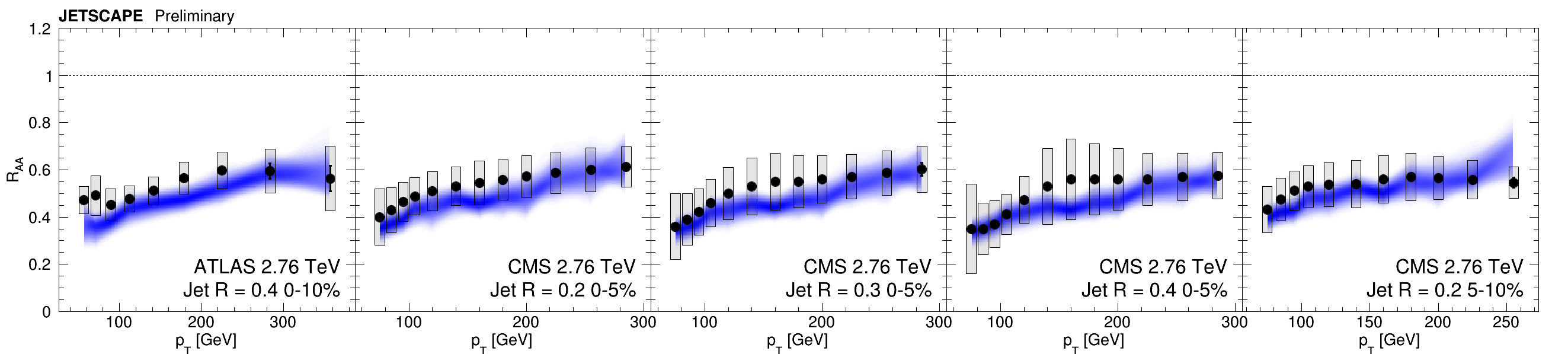}
    \includegraphics[width=0.8\textwidth]{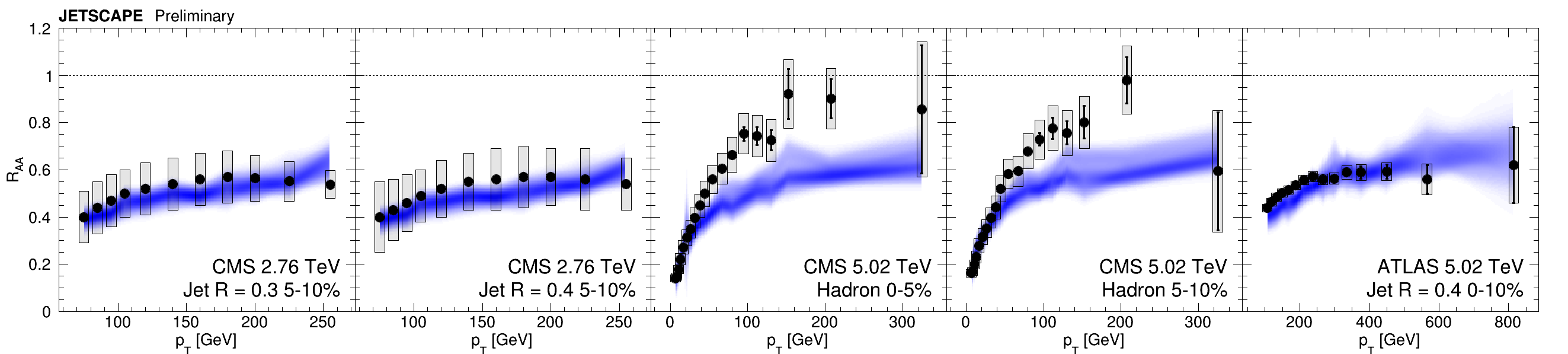}
    \includegraphics[width=0.8\textwidth]{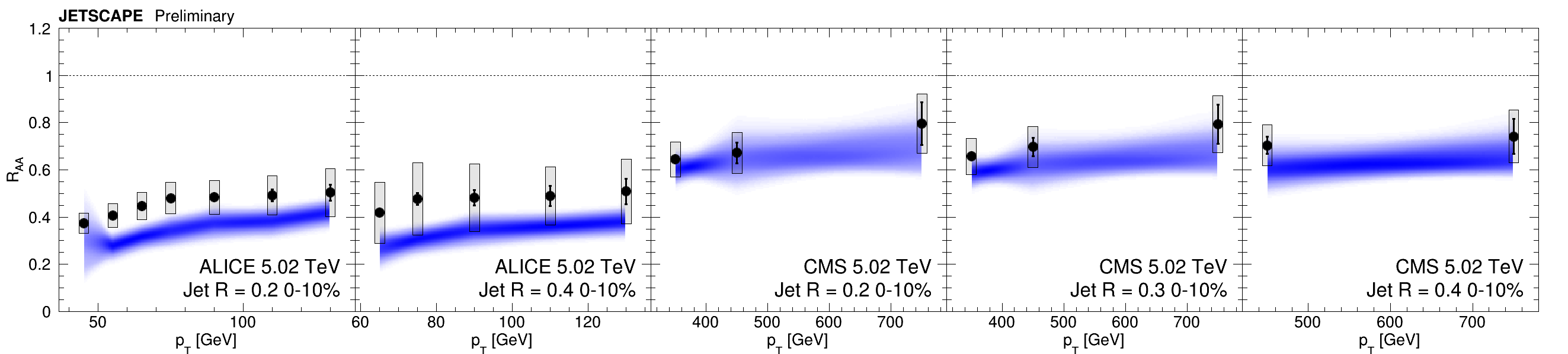}
    \caption{Preliminary posterior distribution of the calibrated model compared to a selection of inclusive jet and charged hadron $\Raa{}$ data.
    The data are in black and the sampled posterior is in blue.}
    \label{fig:hadronAndJetRaaPosterior}
\end{figure}

Figure~\ref{fig:hadronAndJetRaaPosterior} shows the preliminary observable posterior distribution for a multi-stage model using MATTER+LBT. These results only utilize a subset of the JETSCAPE simulations. The model is able to describe the data fairly well overall, although there are some regions of tension. Detailed description of this work-in-progress analysis is available in Ref.~\cite{Ehlers:2022ulm}, with full results following in a separate publication.

\begin{figure}[p]
    \centering
    \includegraphics[width=0.9\textwidth]{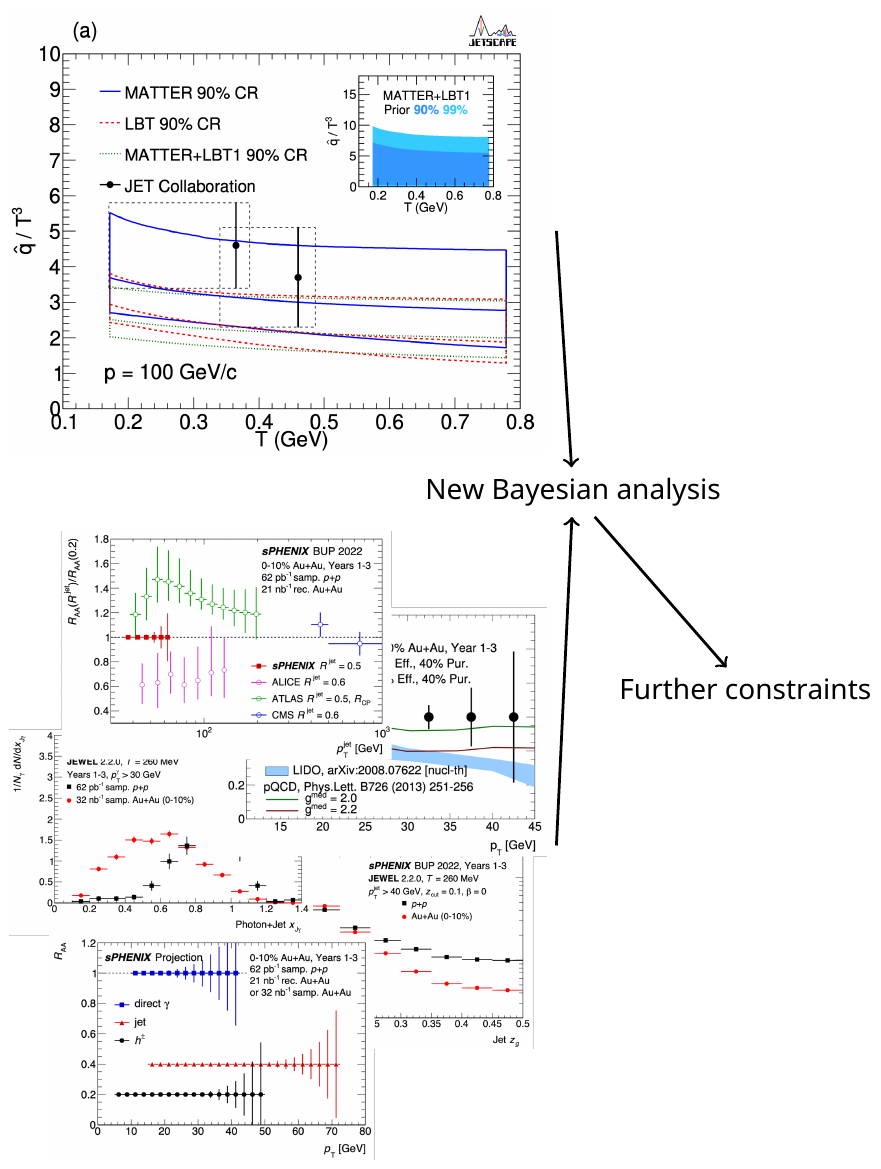}
    \caption{A schematic description of the application of Bayesian sensitivity quantification to upcoming sPHENIX measurements at RHIC.
    The combination of an existing Bayesian analysis with pseudo-data simulations of future data enables quantification of the impact of proposed measurements.}
    \label{fig:bayesianSensitivityQuantificationCartoon}
\end{figure}

Bayesian inference can also be used for sensitivity studies, to assess the impact of future measurements.
This approach is referred to as Bayesian sensitivity quantification, and is a concept that is well established in other fields, including in neutrino physics~\cite{AshtariEsfahani:2020bfp}. In heavy-ion physics it has been applied in the soft sector for oxygen-oxygen collisions~\cite{Nijs:2021clz}.
These studies are implemented by calibrating a model using Bayesian inference, generating pseudo-data by running new simulations according to the extracted parameter posterior distribution, and then including the pseudo-data in a new Bayesian inference analysis.
By varying the included observables and their precision, their future impact can be quantified.
Figure~\ref{fig:bayesianSensitivityQuantificationCartoon} shows a cartoon of such an approach.
This technique can be applied to jet quenching measurements, utilizing the full results from the new JETSCAPE Bayesian analysis reported here to identify which observables will have the largest impact on physics parameter extraction.
This information can help to prioritize measurement strategies, thereby maximizing the impact of new data from sPHENIX in the upcoming RHIC runs.

\section{Heavy flavor quark probes of the QGP}

A second major scientific motivation for the sPHENIX experiment is its comprehensive program of heavy-flavor physics measurements. In the open charm sector, the precision vertex detectors will provide the capability for measurements of the production and azimuthal correlation of $D$ meson resonances (including $v_1$), $D^0$-tagged jets, and the relative yields of charmed baryons such as the $\Lambda_c$. In the open beauty sector, this includes measurements of fully reconstructed $B$ hadrons through the intermediate identification of non-prompt $D$ mesons and the algorithmic tagging $b$-jets. In the quarkonium sector, sPHENIX will have sufficient momentum resolution to separate the three Upsilon states, as well as provide a large-statistics sample of high-$p_\mathrm{T}$ $J/\psi$'s for study. Finally, the streaming readout capability of the trackers will enable high-statistics measurements of key topics in small systems, such as the presence or absence of collective motion for heavy flavor in these systems at RHIC energies. A number of predictions for heavy-flavor hadron and quarkonia probes of the QGP were presented at the workshop and are summarized below.

\subsection{DREENA-A predictions for open heavy flavor in sPHENIX}

The main idea behind QGP tomography is that when high-$p_\mathrm{T}$ particles transverse QGP, they lose energy. This energy loss is sensitive to QGP properties, and thus comprehensive comparisons between high-$p_\mathrm{T}$ theory and data can be used to infer some bulk QGP properties. However, to implement this idea, it is crucial to have a reliable high-$p_\mathrm{T}$ parton energy loss model. This contribution describes a dynamical energy loss formalism~\citep{Djordjevic:2009cr,Djordjevic:2008iz,Djordjevic:2006tw}, developed with the following unique features: i) The formalism takes into account finite size, finite temperature QCD medium consisting of dynamical (that is, moving) partons, contrary to the widely used static scattering approximation and/or medium models with vacuum-like propagators. ii) The calculations are based on the finite temperature generalized Hard-Thermal-Loop approach, in which the infrared divergences are naturally regulated, so there are no artificial cutoffs. Non-perturbative effects related to screening of the chromo-magnetic and chromo-electric fields are also included, as well as a running coupling~\cite{Djordjevic:2011dd,Djordjevic:2013xoa}. iii) Radiative and collisional energy losses are calculated under the same theoretical framework, applicable to both light and heavy ﬂavor. iv) Importantly, there are no fitting parameters in the model, and the temperature ($T$) is a natural variable in the model.

The developed framework further needs to include a full, arbitrary, medium evolution in the dynamical energy loss. All energy loss properties must be preserved, without additional simplifications in the numerical procedure, since all of them are necessary to accurately explain the data. The computational procedure must also be efficient (time-wise) to generate a comprehensive set of light and heavy flavor suppression predictions through the same numerical framework and parameter set. Such predictions can then be compared with the comprehensive set of available experimental data (for different probes, collision systems, energies, centralities), if needed iteratively for different combinations of QGP medium parameters, to extract medium properties consistent with low and high-$p_\mathrm{T}$ theory and data. These goals were achieved by developing a fully optimized DREENA-A framework~\cite{Zigic:2021rku}, where DREENA stands for Dynamical Radiative and Elastic ENergy loss Approach. ``A'' stands for Adaptive temperature profile, meaning that arbitrary medium evolution that can be used as an input.

\begin{figure}[t]
\begin{center}
\includegraphics[width=0.95\textwidth]{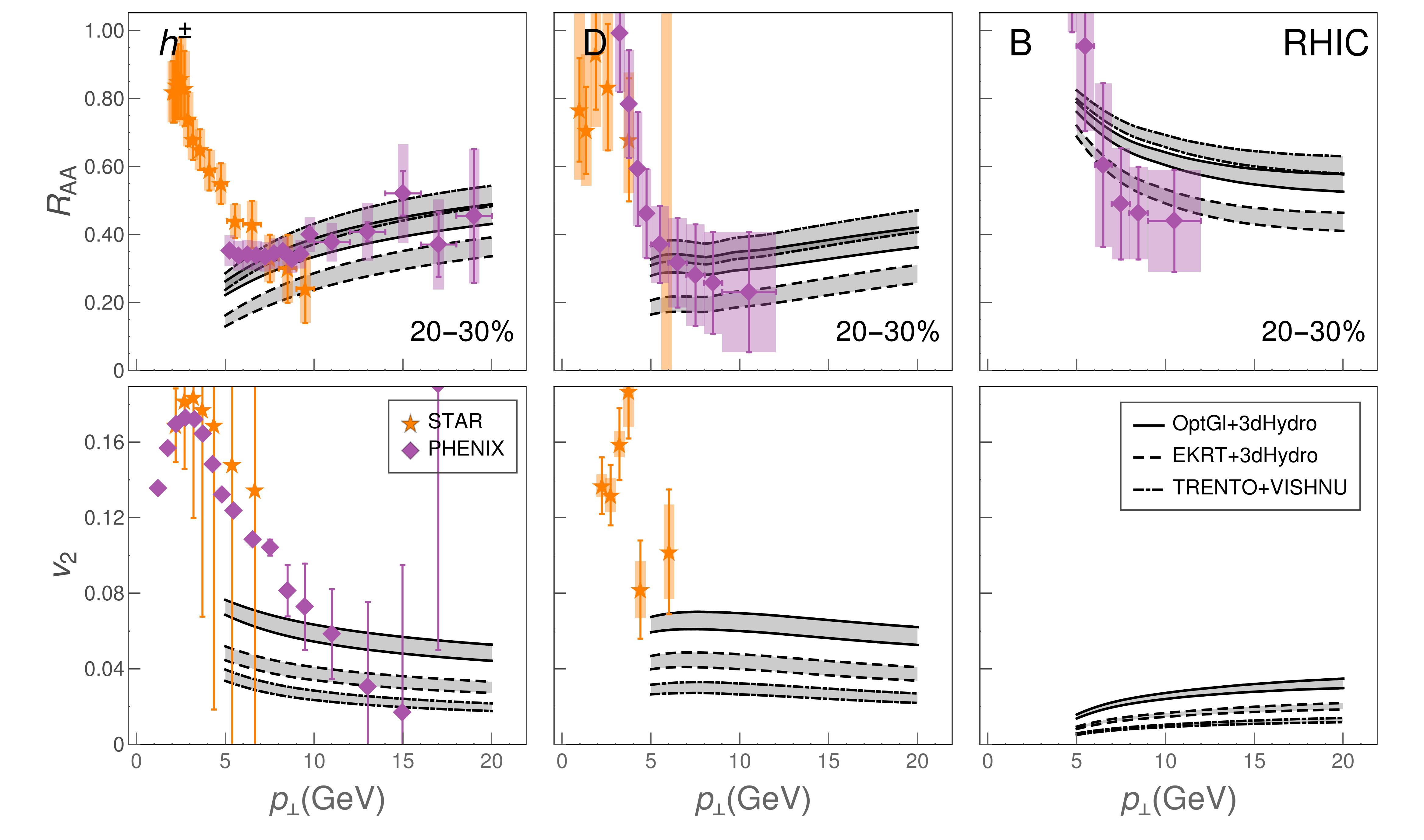}
\end{center}
\caption{ DREENA-A $R_{AA}$ (top panels) and $v_2$ (bottom panels)
  predictions for 200 GeV Au+Au collisions at RHIC are
  generated for three different QGP evolution models (indicated in
  the legend). Predictions are generated for 20-30\% centrality region for charged hadron (left), D meson (middle) and B meson (right), and compared with the available experimental data from RHIC. Figure adapted from~\cite{Zigic:2021rku}.}\label{RHIC}
\end{figure}

The next question is if one can indeed expect different $T$ profiles to lead to differences in high-$p_\mathrm{T}$ observables. To address this question, representative evolutions were generated (see~\cite{Zigic:2021rku} for more details): i) Optical Glauber initialization at initial time $\tau_0 = 1.0$ fm, without initial transverse flow, followed by 3+1D viscous fluid expansion. ii) EKRT initialisation with $\tau_0 = 0.2$ fm, also followed 3+1D viscous fluid dynamics. iii) T$_\mathrm{R}$ENTo
initialisation evolved by free streaming until $\tau_0 = 1.16$ fm, followed by VISH2+1 viscous fluid dynamics. While they all agree with low-$p_\mathrm{T}$, they lead to quite different $T$ profile evolutions with time, as discussed in ~\cite{Zigic:2021rku}. Can high-$p_\mathrm{T}$ data further constrain these evolutions? To address this, in Fig.~\ref{RHIC}, these profiles were used as an input to the DREENA-A to generate high-$p_\mathrm{T}$ $R_{AA}$ and $v_2$ predictions for charged hadrons, D and B mesons at RHIC. Both the $R_{AA}$ and the $v_2$ calculations show notable differences for all types of flavor. Consequently, the DREENA-A framework can clearly differentiate between $T$ profiles by corresponding differences in
high-$p_\mathrm{T}$ observables. Furthermore, heavy-flavor hadrons show higher sensitivity than light flavor, making them even better suited for exploring the bulk QGP parameters with high-$p_\mathrm{T}$ data. With the expected availability of precision data from the upcoming high-luminosity experiments at sPHENIX, the DREENA-A framework will provide an exciting new opportunity~\cite{Stojku:2020wkh,Stojku:2021yow} for exploring the bulk QGP properties at heavy ion collisions.

\subsection{Heavy flavor jets in sPHENIX with LIDO}

The LIDO model \cite{Ke:2018tsh, Ke:2018jem} is a linearized partonic transport model for jet and heavy-flavor transport in the quark-gluon plasma. The jet parton undergoes multiple collisions with QGP constituents, treated in a small-angle diffusion plus large-angle perturbative scattering approach \cite{Ke:2018tsh}. Medium-induced parton branchings (including $g\rightarrow g+g$, $q\rightarrow q+g$, $g\rightarrow q+\bar{q}$, with $q=u,d,s,c,b$) are included both in the incoherent and the deep LPM limits. Especially, the LPM suppression is implemented dynamically in the simulation so that the finite-size and expanding medium features of the radiation pattern are reproduced \cite{Ke:2018jem}.

To study jets, a matching routine is developed to transit the Pythia vacuum parton shower generation and the LIDO on-shell parton transport. A simple medium response to the hard-parton energy deposition is instrumented to guarantee the energy-momentum conservation in the simulation of the jet event. The Pythia8+LIDO framework with a 2+1D hydrodynamic simulation of heavy-ion collisions has been calibrated to the single-inclusive light and charm meson nuclear modification factor $R_{AA}$ and the $R=0.4$ jet $R_{AA}$ at both RHIC and LHC energies in central $A$-$A$ collisions \cite{Ke:2020clc}.

\begin{figure}
    \centering
    \includegraphics[width=.9\textwidth]{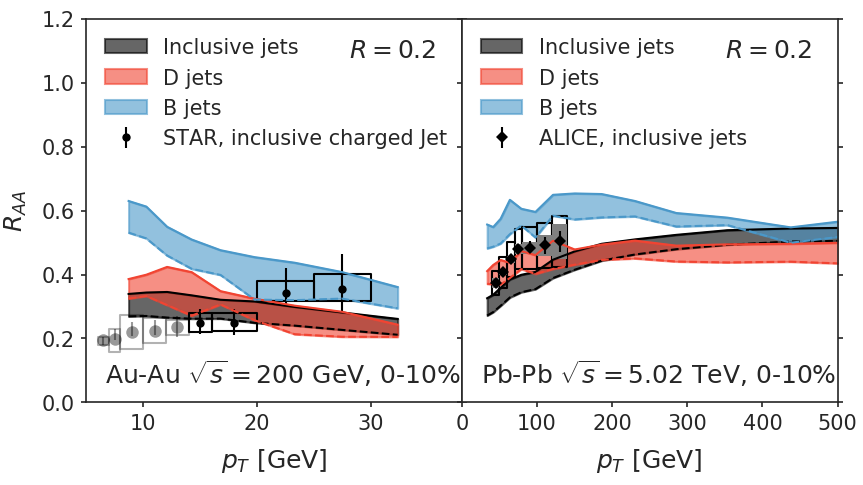}
    \caption{Flavor dependence of full jet suppression. Left: projection in central Au+Au collisions at $\sqrt{s}$=200 GeV. Right: in central Pb+Pb collisions at $\sqrt{s}$=5.02 TeV. The flavor hierarchy comes from both the dead-cone effect and the respective contributions from $Q\rightarrow$ HF jets (dominated at low $p_T^{\rm jet}$) and $g\rightarrow$ HF jets. }
    \label{fig:lido:HFjetRaa}
\end{figure}

\begin{figure}
    \centering
    \includegraphics[width=.9\textwidth]{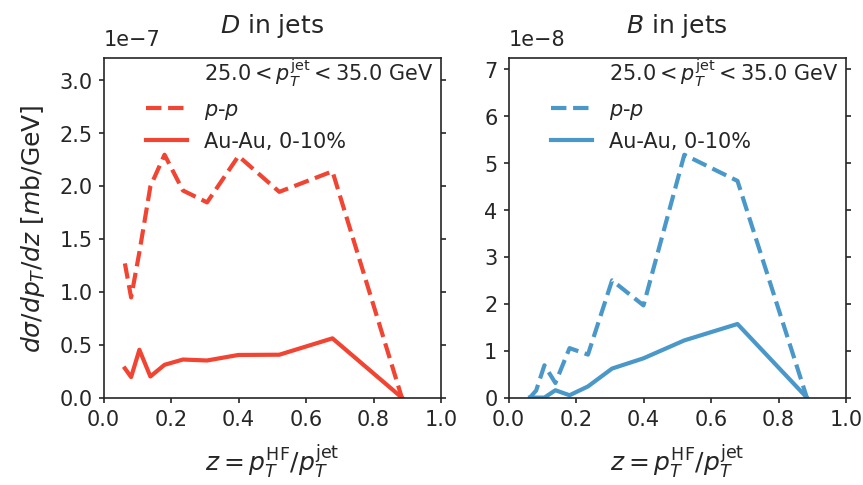}
    \caption{$D$ (left) and $B$ (right) meson fragmentation functions in jets. 
    The Peterson fragmentation functions are used in both cases for heavy flavor fragmentation in the vacuum. }
    \label{fig:lido:FF}
\end{figure}

The framework is recently applied to study heavy-flavor (HF) jet  quenching \cite{Ke:2020nsm}: the flavor-dependent jet $R_{AA}$ (Fig. \ref{fig:lido:HFjetRaa}) and the heavy-flavor-in-jet fragmentation function (Fig. \ref{fig:lido:FF}) at jet $p_T$ relevant for the sPHENIX experiment. Medium modifications of heavy-flavor jets are controlled by not only the mass dependence of in-medium parton branchings but also the competing channels of $Q\rightarrow $HF jets and $g\rightarrow $ HF jets. HF jet production at RHIC energy and low $p_T^{\rm jet}$ is dominated by $Q\rightarrow$ HF jets fragmentation. It is very different from the situation at the LHC energy, providing an independent constraint to disentangle quark/gluon contributions and flavor-dependent jet energy loss. Additionally, HF-in-jet measurements directly allows one to extract both the vacuum and in-medium fragmentation functions of heavy quarks and provide a more differential test of the heavy-flavor dynamics in the medium. 

\subsection{KSU and QTraj predictions for Upsilon suppression in sPHENIX}


One of the key motivations used for the construction of sPHENIX was to study the suppression of bottomonium states, e.g., the $\Upsilon(1S)$ and $\Upsilon(2S)$, in RHIC energy heavy-ion collisions relative to their production in $pp$ collisions.  The strong suppression of such states provides crucial information about whether or not a quark-gluon plasma has been created.  In addition, such studies can provide constraints on the initial temperatures generated during the collisions and in-medium bottomonium transport properties that can be checked against first principles non-relativistic QCD and lattice calculations.  Due to the largeness of the bottom quark mass, such quarks are created only in the initial hard scatterings and it is not expected that there will be a significant regeneration/recombination effect due to the smallness of both open- and closed-bottom production cross-sections.  Additionally, due to their large mass, theoretical calculations which rely on non-relativistic effective field theories (EFTs) are in much better control for describing the physics of bottomonium than for charmonium.

Related to this, in the last decade there has been a significant advance in the understanding of bottomonium dynamics in the QGP through the use of EFTs and open quantum system methods \cite{Brambilla:2016wgg,Brambilla:2017zei}.  Most recently, these methods have been applied to make phenomenological predictions for bottomonium suppression and flow at LHC enerigies~\cite{Brambilla:2020qwo,Brambilla:2021wkt,Brambilla:2022ynh}.  In practice, one solves for the evolution of the heavy-quarkonium reduced density matrix using a quantum master equation, which at high temperatures can be cast in the form of the Lindblad equation \cite{Lindblad:1975ef,Gorini:1975nb}.  This equation can be efficiently solved using a realistic 3+1D dissipative hydrodynamic background \cite{Alqahtani:2020paa} by making use of an algorithm called the ``quantum trajectories algorithm'', which allows for massive parallel computations.  This algorithm has been implemented for heavy-quarkonium dynamics in a publicly available package called QTraj~\cite{Omar:2021kra}.  

\begin{figure}[t]
\centerline{
\includegraphics[width=0.49\linewidth]{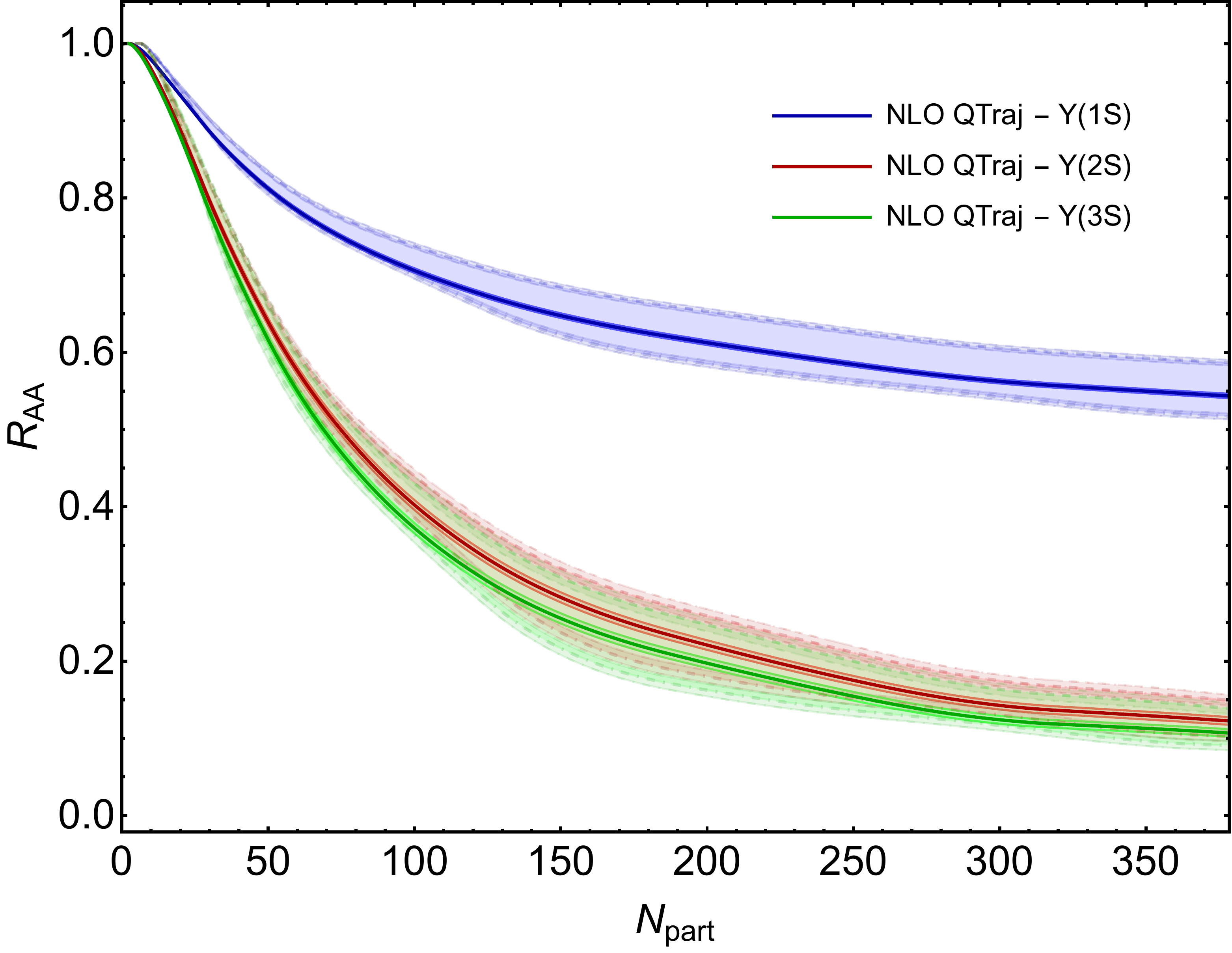}
\hspace{2mm}
\includegraphics[width=0.49\linewidth]{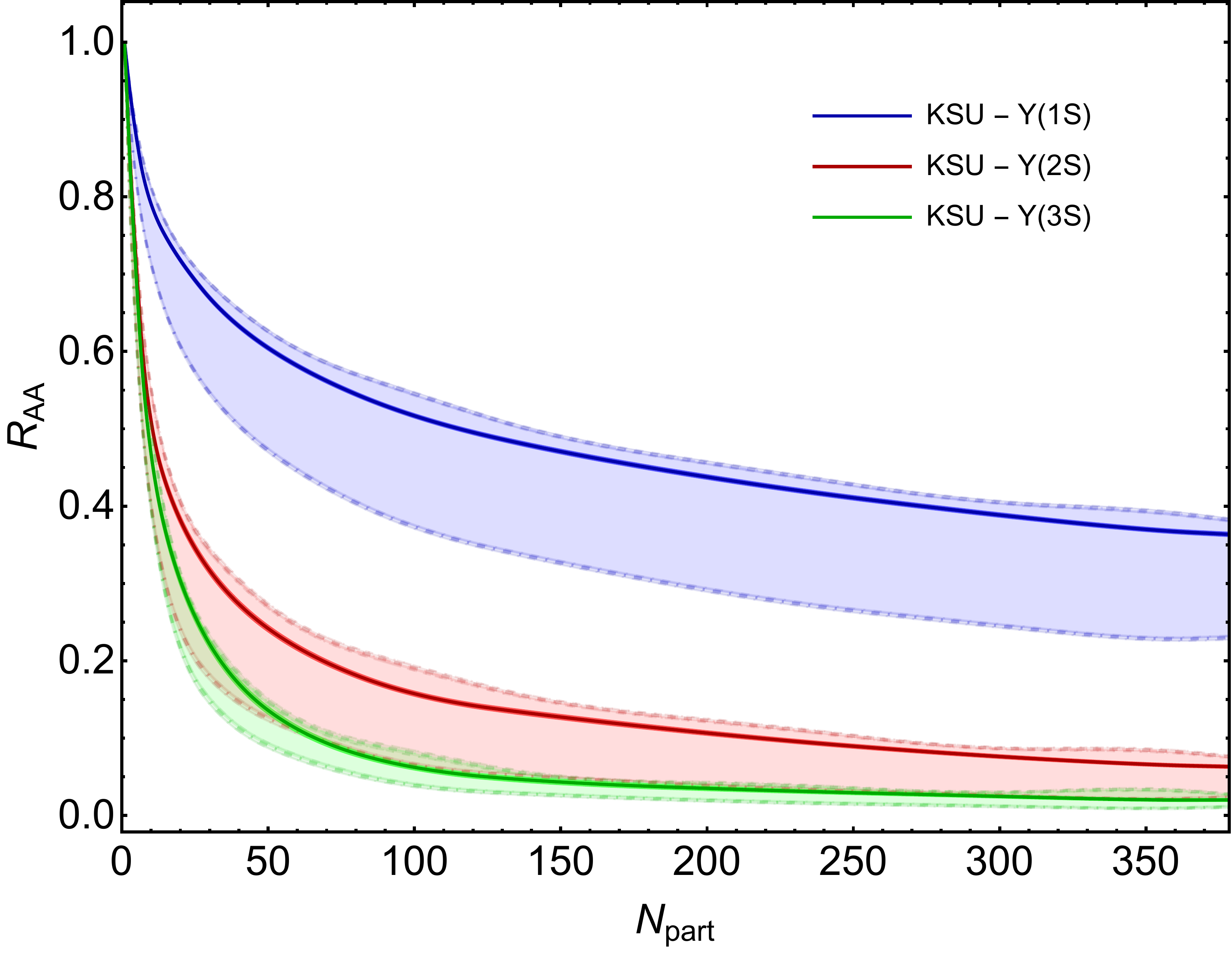}
}
\caption{Predictions for bottomonium suppression as a function of $N_{\rm part}$ at RHIC collision energies.  The left panel shows the result obtained using Coulomb-like potentials emerging from an EFT calculation~\cite{Brambilla:2020qwo,Brambilla:2021wkt,Brambilla:2022ynh} and the right panel shows the results emerging from the use of a phenomenological potential which reduces to a Cornell potential in the $T=0$ limit~\cite{Islam:2020gdv,Islam:2020bnp}.  The bands in both figures are described in the text. }
\label{fig:strickland}
\end{figure}

Although the EFT formalism used is most trustworthy at high temperatures, it can be used to make predictions at the lower temperatures generated at RHIC. For this purpose, once again one can use 3+1D dissipative hydrodynamics codes that have been tuned to agree with the experimentally observed identified soft-hadron production and flow \cite{Almaalol:2018gjh}.  The left panel of Fig.~\ref{fig:strickland} presents the predictions of the EFT approach using QTraj at RHIC 200 GeV collision energy.  For comparison, in the right panel of Fig.~\ref{fig:strickland}, a different underlying potential is used that, instead of being Coulomb-like, includes a non-trivial long range part which allows it to reduce to a Cornell potential in the $T=0$ limit~\cite{Islam:2020gdv,Islam:2020bnp}. This second model is labelled as the Kent State University (KSU) model due to its origins.  In the case of the QTraj predictions, the bands are obtained by varying the heavy-quarkonium transport coefficients $\kappa$ in the range suggested by lattice calculations and, in the case of the KSU predictions, the bands are obtained by varying the effective Debye mass around its leading-order QCD value.

\subsection{Modeling quarkonium suppression in sPHENIX}

This contribution describes a simple framework to include initial shadowing and medium effects in the computation of quarkonium related observables in heavy ion collisions \cite{Escobedo:2021ifp}. The approach presented here is particularly suitable when the survival probability of a quarkonium state that traverses the plasma can be expressed as a simple analytic function of the initial temperature. In order to compute the nuclear modification factor $R_{AA}$ in this situation, one needs to answer the following questions: what is the initial temperature for a given collision type, how does it depend on the transverse position, and how does the probability to create a quarkonium state depend on the transverse position. 

The following computation is based on the shadowing model discussed in \cite{Capella:2011vi}. The Glauber model can be obtained assuming that nucleons interact exchanging pomerons. This framework can be modified to include shadowing effects by introducing a triple pomeron vertex. The initial temperature distribution is computed by assuming that the temperature scales with the energy density to the power of $1/4^\mathrm{th}$ and that the initial energy density scales with the number of pions. The proportionality constant between the energy density and the pion production is not known, however, this information is not needed to compute the ratio of energy densities between two different points in transverse space and/or impact parameter. It was found that the initial temperature scales approximately with the density of participants (or wounded nucleons) at a given point, both for the conditions at RHIC and the LHC. Using the same shadowing model, the initial distribution of quarkonium is computed, which scales approximately with the density of binary collisions. 

\begin{figure}[t]
\includegraphics[width=.48\textwidth]{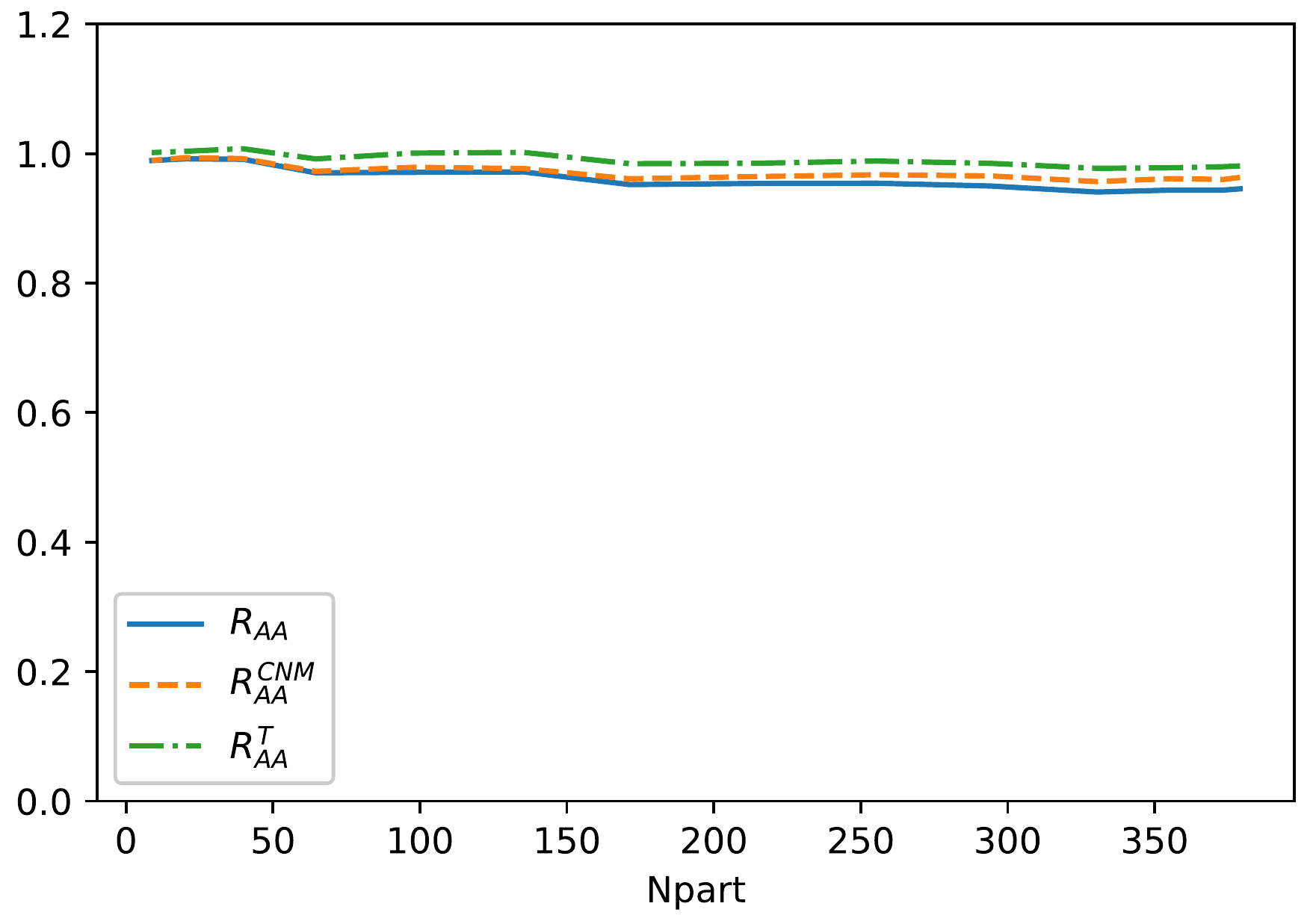}
\includegraphics[width=.48\textwidth]{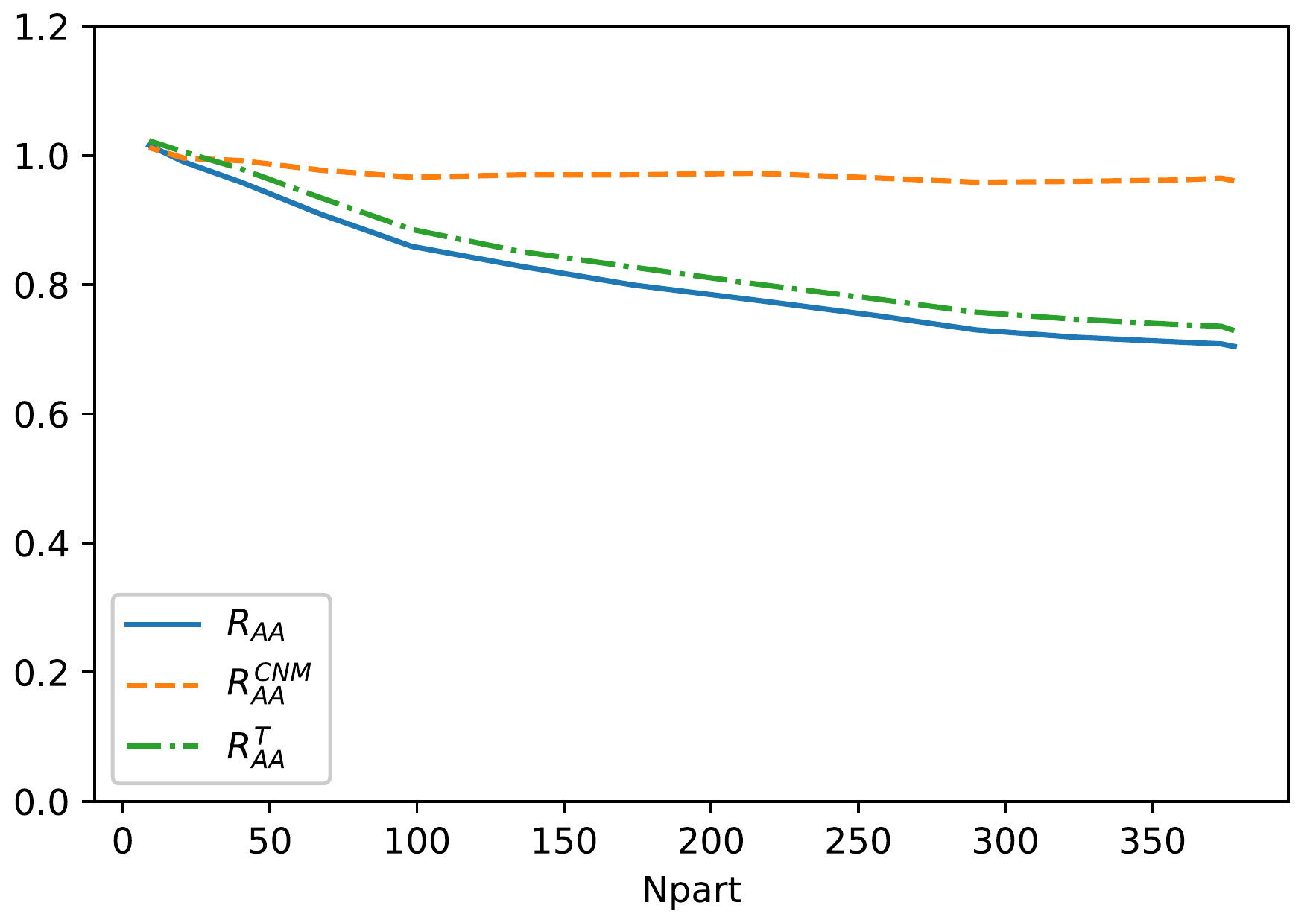}
\caption{Predictions for the $R_\mathrm{AA}$ vs $N_\mathrm{part}$ of $\Upsilon(1S)$ (left) and $\Upsilon(2S)$ (right) in sPHENIX at RHIC, showing the cold nuclear matter (orange) and thermal (green) contributions to the total $R_\mathrm{AA}$ (blue). }
\label{fig:raa}
\end{figure}

In order to compute the $R_{AA}$, the survival probability obtained in Ref.~\cite{Blaizot:2021xqa} is applied. The approach of \cite{Blaizot:2021xqa} is valid when thermal effects are a perturbation, in other words, well below the melting temperature. The model employs lattice QCD data on the static potential to obtain the binding energy and the wave function. These two quantities are then used to determine the decay width applying the Hard Thermal Loop approach. The focus of this model is on the finite energy gap that exists between singlets and octets. This gap strongly suppresses the decay width at low temperatures. The results are shown in Fig.~\ref{fig:raa} and feature a very mild medium effect on the $R_{AA}$ of $\Upsilon(1S)$ at sPHENIX. 
However, one should take into that this computation refers to the suppression suffered by the direct state and does not take into account the feed-down effects.

In fact, in order to compare with experimental inclusive data, it is mandatory to take into account the feed-down contributions.
The feed-down fractions for the $\Upsilon$(1S) can be estimated as: 70\% of direct $\Upsilon$(1S), 8\% from $\Upsilon$(2S) decay, 1\% from $\Upsilon$(3S),
15\% from $\chi_{\rm B1}$, 5\% from $\chi_{\rm B2}$ and 1\% from $\chi_{\rm B3}$, 
while for the $\Upsilon$(2S) the different contributions would be: 63\% direct $\Upsilon$(2S), 4\% of $\Upsilon$(3S), 30\% of $\chi_{\rm B2}$ and 3\% of  $\chi_{\rm B3}$  \cite{Andronic:2015wma}. Note also that for the  $\Upsilon$(3S), 40\% of the contribution will come from decays of  $\chi_{\rm B3}$. Those estimations are based on data measured by the LHCb Collaboration \cite{LHCb:2014ngh} at low $p_T$. The contribution of the excited states could be even higher according to CDF Collaboration measurements at $p_T > 8$ GeV \cite{CDF:1999fhr}.

Thus, in order to make predictions for the upcoming sPHENIX data, the following approach was developed.

First, consider that within this model $\Upsilon(3S)$ and the excited $\chi_{\rm B2}$ and $\chi_{\rm B3}$  states are totally dissociated in Au+Au collision at RHIC energies. 
In this case, the nuclear modification factor for inclusive $\Upsilon(2S)$ production will achieve a value around 0.4 for the most central collisions. 
Consequently,  the estimated value for $\Upsilon(1S)$ will vary between 0.7 (according to LHCb feed-down estimations) and 0.5 (according to CDF feed-down estimations) at high centrality.

This contribution explores the phenomenological consequences of taking into account the energy gap between singlet and octet states when computing the decay width of an upsilon bound state in a medium. Moreover, initial shadowing effects are also included. This energy gap and the shadowing corrections induce sizeable effects for the conditions in Au+Au collisions at RHIC and should be taken into account in phenomenological studies.

\section{Bulk probes of the QGP}

Although originally motivated as an experiment for rare probes, the large tracking acceptance and high rate capabilities of sPHENIX are highly suited to measurements of soft, collective phenomena such as correlations and fluctuations of the flowing QGP medium. Additionally, characterizing the QGP bulk provides important information for dealing with the backgrounds present in jet and heavy flavor measurements. Below is a summary of one such physics opportunity presented at the workshop.

\subsection{Non-Gaussian flow fluctuations at sPHENIX}


The sPHENIX program will open a new window onto the initial condition of the QGP at RHIC via precision measurements of the event-by-event distribution of the anisotropic flow coefficients, $v_n$. In particular, sPHENIX will permit us to access the non-Gaussianities of these distributions, which are of fundamental interest as they emerge from the non-Gaussian properties of the fluctuating energy density field characterizing the QGP initial condition on an event-by-event basis \cite{Floerchinger:2014fta,Gronqvist:2016hym,Bhalerao:2019fzp}. While precision measurements of non-Gaussian flow fluctuations have been achieved in Pb+Pb collisions at the Large Hadron Collider \cite{ATLAS:2014qxy,CMS:2017glf,ALICE:2018rtz,ATLAS:2019peb}, two key results are currently missing in the Au+Au data sets at RHIC. The first involves the skewness of the fluctuations of elliptic flow, $v_2$, which emerges in the fine splitting (of order 1\%) between the fourth-order cumulant, $v_2\{4\}$, and the sixth-order cumulant, $v_2\{6\}$, of $v_2$ fluctuations. Linear response to the initial QGP ellipticity, $v_2 \propto \varepsilon_2$ \cite{Teaney:2010vd,Niemi:2012aj}, yields a negatively-skewed distribution of $v_2$ in off-central collisions ($>10\%$) due to the bound $\varepsilon_2<1$ \cite{Giacalone:2016eyu}. Precision measurements of the ratio $v_2\{6\} / v_2\{4\}$ offer, thus, a precision tool to scrutinize the hydrodynamic response of the QGP. 

The second important result that will be accessible to sPHENIX is the kurtosis of the fluctuations of $v_3$, quantified by the four-particle cumulant $c_3\{4\}$ \cite{Abbasi:2017ajp,Bhalerao:2018anl,Bhalerao:2019fzp}. This quantity also emerges from the response to an initial triangularity, $v_3 \propto \varepsilon_3$ \cite{Alver:2010gr,Teaney:2010vd,Niemi:2012aj}, whose distribution is non-Gaussian (with a negative kurtosis), potentially due to the positivity constraint on the local energy density field, $e>0$, shaping event-to-event the initial condition of the QGP \cite{Bhalerao:2019fzp}. While measurements of $c_3\{4\}$ have been reported by the PHENIX collaboration \cite{PHENIX:2018lfu}, these have been performed only via correlations of particles present in $1<|\eta|<3$ that show considerable effects of poorly-understood longitudinal fluctuations. At midrapidity, where the response to the initial geometry is more robustly understood, the kurtosis obtained in the existing data sets is compatible with zero \cite{STAR:2013qio}. 

In this section, the linear response, $v_n \propto \varepsilon_n$, is exploited to provide solid baselines for non-Gaussian flow fluctuations at sPHENIX. Ratios of cumulants are constructed that isolate the impact of the skewness of $v_2$ fluctuations and of the kurtosis of $v_3$ fluctuations, namely,
\begin{equation}
    \frac{v_2\{6\}}{v_2\{4\}}\approx\frac{\varepsilon_2\{6\}}{\varepsilon_2\{4\}}, \hspace{50pt} \frac{c_3\{4\}}{c_3\{2\}^2 }  \approx \frac{\varepsilon_3\{4\}^4}{\varepsilon_3\{2\}^4}.
\end{equation}
The anisotropies $\varepsilon_n$ are obtained with the \trento{} model of initial conditions \cite{Moreland:2014oya}, tuned to reproduce LHC data as in Ref.~\cite{Bally:2021qys}, where the centrality is evaluated from the entropy of the collisions. No parameters related to the geometry of the QGP are changed, most notably the size of nucleons, when moving from LHC and RHIC collisions, to provide a baseline of results that does not contain any energy-dependent effects. Results are shown up to 40\% centrality, a range where linear hydrodynamic response is an excellent approximation.

\begin{figure*}[t]
    \centering
    \includegraphics[width=0.98\linewidth]{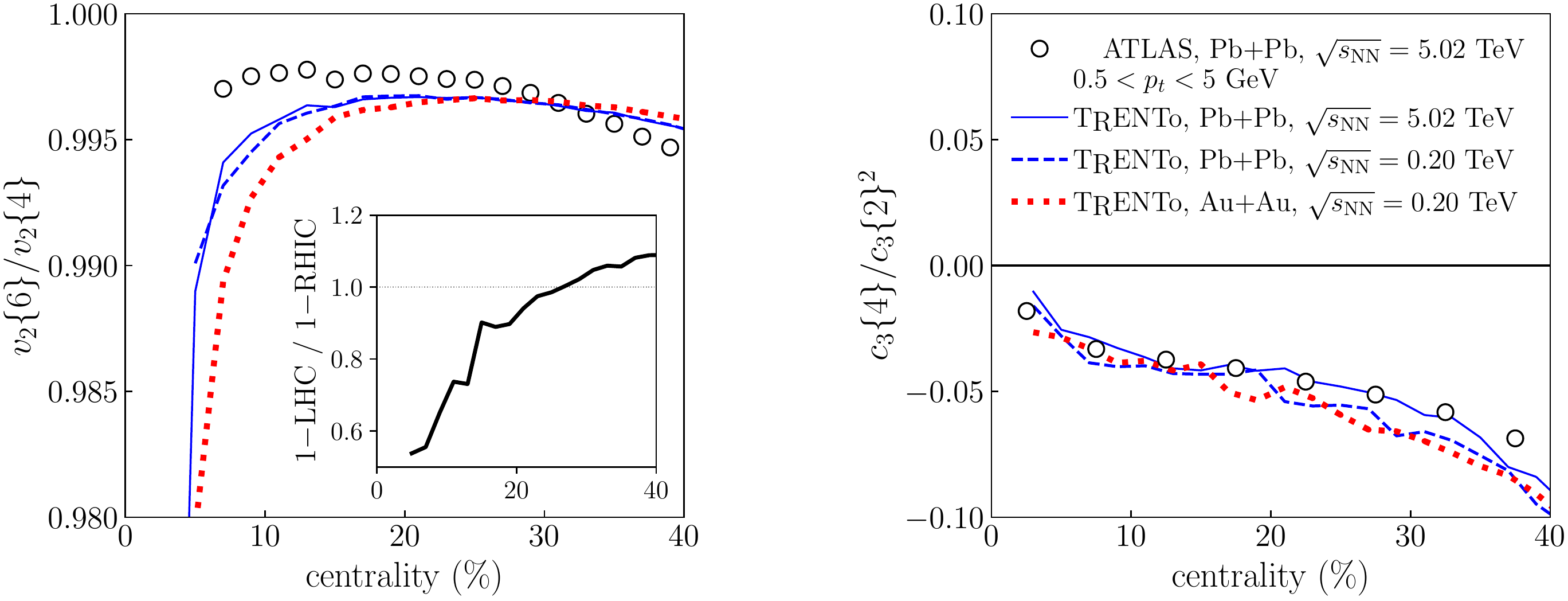}
    \caption{Non-Gaussian fluctuations of $v_2$ and $v_3$ in heavy-ion collisions. Left: $v_2\{6\}/v_2\{4\}$. Right: $c_3\{4\}/c_3\{2\}^2$. Experimental data (symbols) are measurements by the ATLAS Collaboration in 5.02 TeV Pb+Pb collisions \cite{ATLAS:2014qxy}. Lines are estimates with the \trento{} model of initial conditions. Solid line: 5.02 TeV Pb+Pb collisions. Dashed line: 200 GeV Pb+Pb collisions. Dotted line: 200 GeV Au+Au collisions. The red dotted lines represent   predictions for sPHENIX.}
    \label{fig:1}
\end{figure*}

The left panel of Fig.~\ref{fig:1} shows predictions for the splitting between $v_2\{4\}$ and $v_2\{6\}$. Results are shown for 200 GeV Au+Au collisions (dotted lines) and 5.02 TeV Pb+Pb collisions (solid lines).  For the latter, the model is observed to be in decent, though not perfect, agreement with data by the ATLAS collaboration \cite{ATLAS:2019peb}. Of interest for this discussion is the prediction relative to the difference between LHC and RHIC, which is shown in the inset panel. Calculations for 200 GeV Pb+Pb collisions (dashed lines) show that this departure does not come from a change in nucleon-nucleon cross section in the \trento{} calculation. Therefore, in the present setup the splitting between Pb+Pb and Au+Au collisions comes likely from the fact that $^{197}$Au has a larger ground-state quadrupole deformation than $^{208}$Pb, which leads to enhanced non-Gaussian $v_2$ fluctuations in central Au+Au collisions \cite{Giacalone:2018apa}. Sizable departures from this basic prediction in future data would point, then, to effects at high energy, most probably from the modification of nucleon structure between RHIC and LHC. Moving on to the fluctuations of $v_3$ in the right panel of Fig.~\ref{fig:1}, the \trento{} estimate of the standardized kurtosis of $v_3$ fluctuations turns out to be in excellent agreement with ATLAS data, strongly supporting the \trento{} Ansatz for the initial energy density fluctuations. The prediction for sPHENIX is shown as a red dotted line. In this case, good agreement is found between the Pb+Pb and Au+Au results.  Significant departures from this prediction will point to energy-dependent effects, most likely coming from the different structure of the colliding nucleons. Precision measurements of the kurtosis from sPHENIX will, hence, pose novel important constraints on the initial condition of the QGP.

\section{QCD with polarized protons and cold nuclei}

In addition to the physics program aimed at exploring the large region of high-temperature QGP created in Au+Au collisions, sPHENIX will explore QCD in two other regimes: the cold but dense regime accessed in $p$+Au collisions, and the vacuum QCD physics in $p$+$p$ collisions which serves as a baseline for interpreting measurements in the other systems. These topics furthermore have a natural connection to physics which will be explored at the future Electron-Ion Collider~\cite{Aschenauer:2017jsk} sited at BNL. Here, two potential measurements are highlighted which are aimed at understanding the multi-dimensional parton distributions in the proton and the parton fragmentation and hadronization process, both utilizing the particular capabilities of sPHENIX.

\subsection{Azimuthal angular correlation in dijet events at sPHENIX}

Recently, the azimuthal angular correlation of dijets has received much attention as a way to probe various multi-dimensional parton distributions inside the nucleon/nucleus. For example, the $\cos 2\phi$ correlation between the dijet relative momentum $p_\mathrm{T}=p_{1\perp}-p_{2\perp}$ and the dijet total momentum $q_\perp=p_{1\perp}+p_{2\perp}$ in the so-called correlation limit $p_\mathrm{T} \gg q_\perp$ has been proposed as a signal of the linearly polarized gluon distribution.

\begin{figure}[t]
  \includegraphics[width=0.49\linewidth]{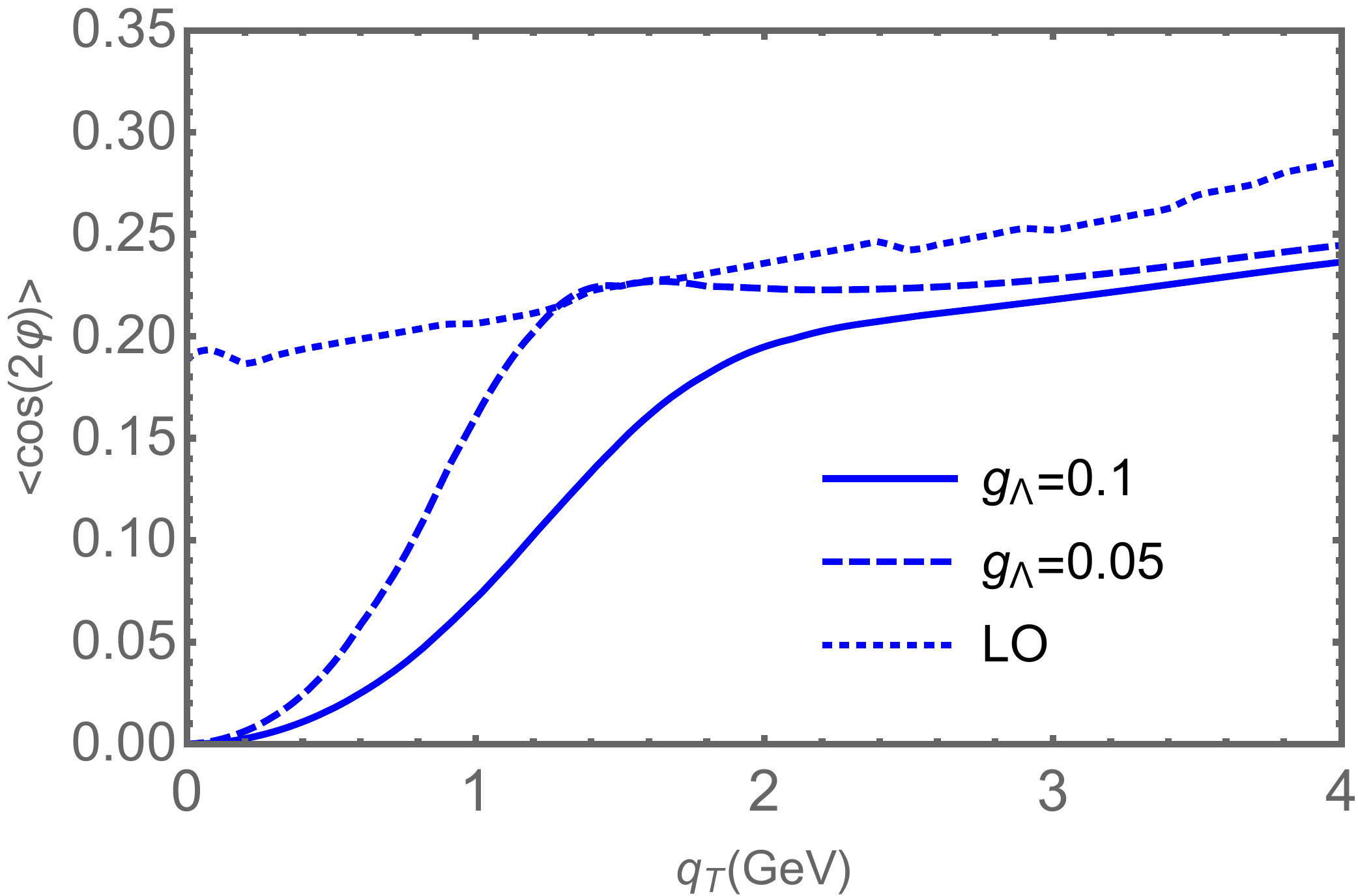}
   \includegraphics[width=0.49\linewidth]{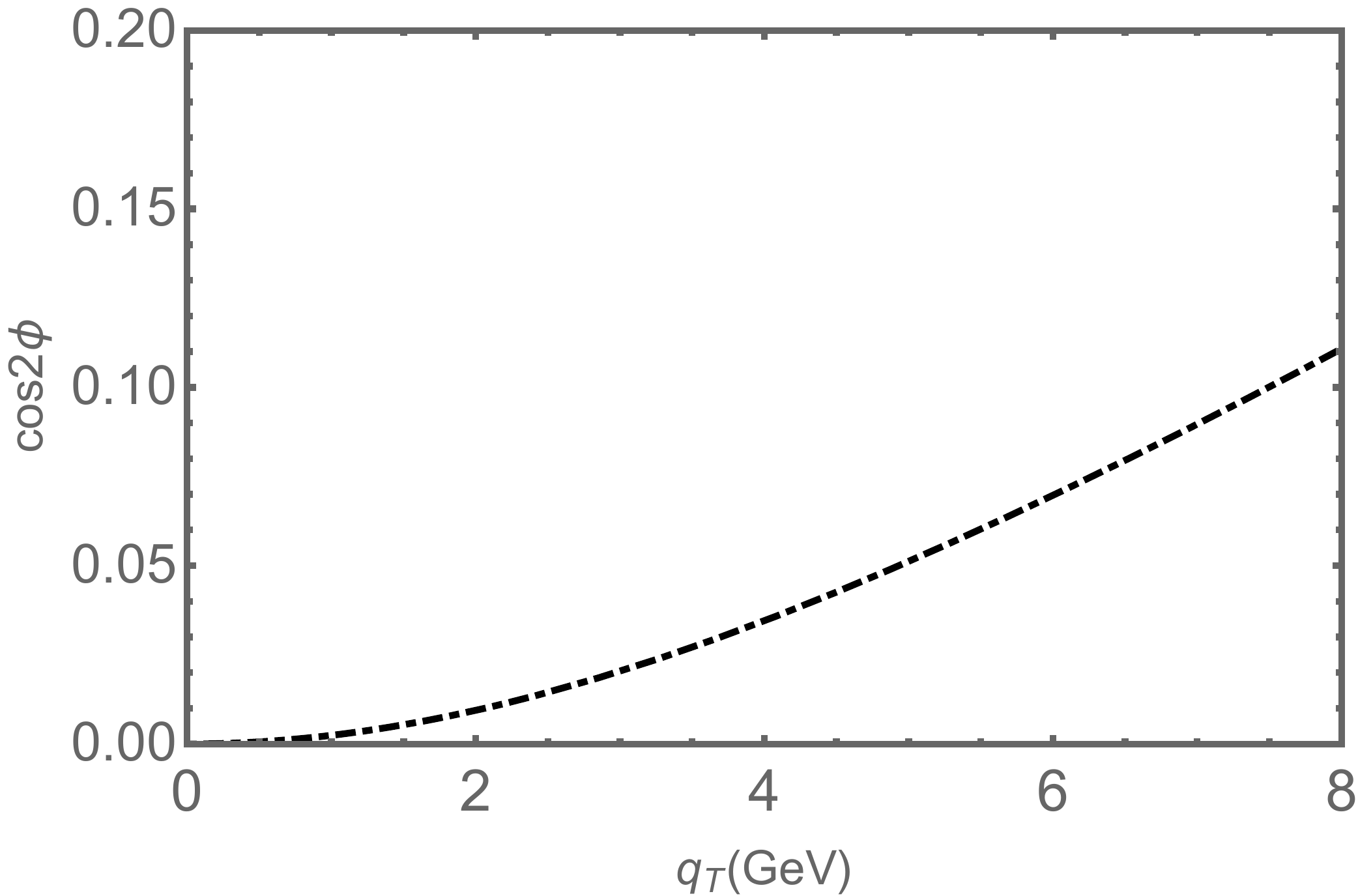}
\caption{$\cos2\phi$ angular correlation in dijet production at midrapidity  in UPCs  (left, $p_\mathrm{T}=10$ GeV) and in inclusive dijet production in $pp$ (right, $p_\mathrm{T}=20$ GeV) at sPHENIX as a  function of the dijet total momentum $q_\perp$.  
}
\label{dijetcol}
\end{figure}

However,  the same angular correlation can also be generated by soft gluon radiations from the incoming and outgoing partons. The resummation of Sudakov logarithms $\sim\ln  p_\mathrm{T}/q_\perp$ associated with angular-dependent soft gluon emissions has been systematically studied in \cite{Hatta:2020bgy,Hatta:2021jcd} in the TMD framework building on an earlier work \cite{Catani:2017tuc}.
Fig.~\ref{dijetcol} shows  predictions for the average $\cos 2\phi$ ($\phi=\phi_{p_\mathrm{T}}-\phi_{q_\perp}$) modulation  at  sPHENIX. The left plot shows diffractive dijet events in UPCs $\gamma p(A) \to jj p'(A')$ and the right plot shows inclusive dijet production in $pp$ collisions $pp \to jjX$.  Since  $p_\mathrm{T}$ is much smaller than at the LHC  (see a recent measurement by the CMS collaboration \cite{CMS:2022lbi}), the asymmetry is more sensitive to the  parameter $g_\Lambda$ that characterizes the nonperturbative Sudakov suppression in the final state jets.  Therefore, future measurements at sPHENIX will be useful to constrain this so far poorly known parameter.  

\subsection{Hadron-in-jet production with sPHENIX}


The study of hadron distributions inside jets has received increasing attention as an effective tool to understand the fragmentation process over the last few years. They allow us to probe the one-dimensional (i.e. collinear) and three-dimensional (i.e. transverse momentum dependent) fragmentation functions~\cite{Kang:2020xyq,Kang:2021ffh}, and thus provide us with a deep insight into the elusive mechanism of hadronization. The unpolarized and transversely polarized proton-proton collisions at the sPHENIX would provide outstanding opportunities for probing and measuring these functions. Below, predictions for two such observables are provided. The first case is a study of the transverse polarization of $\Lambda$ hyperons inside the jets that are produced in unpolarized $pp$ collisions. The second case is a study of the so-called Collins azimuthal asymmetry for charged pions inside the jets in transversely polarized $p^\uparrow p$ collisions. 

\begin{figure}[t]
  \includegraphics[width=.49\linewidth]{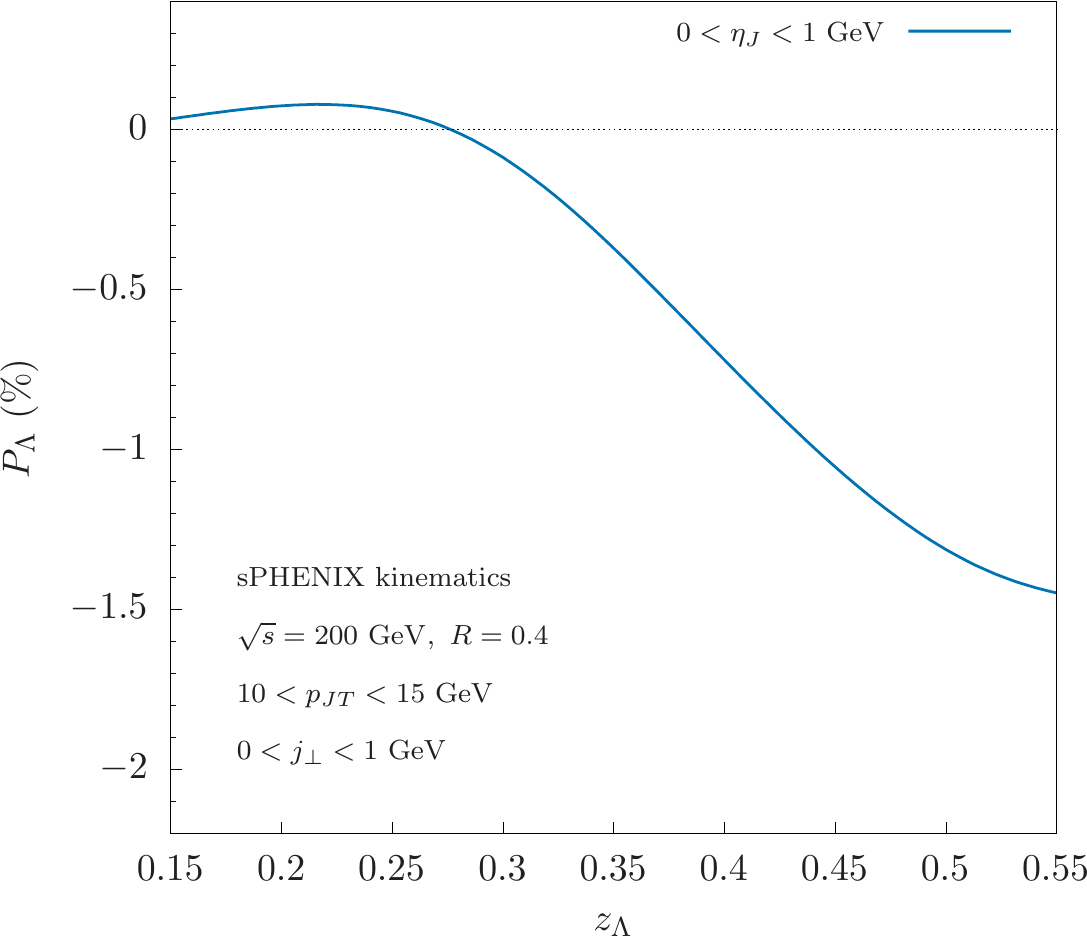}
\includegraphics[width=.49\linewidth]{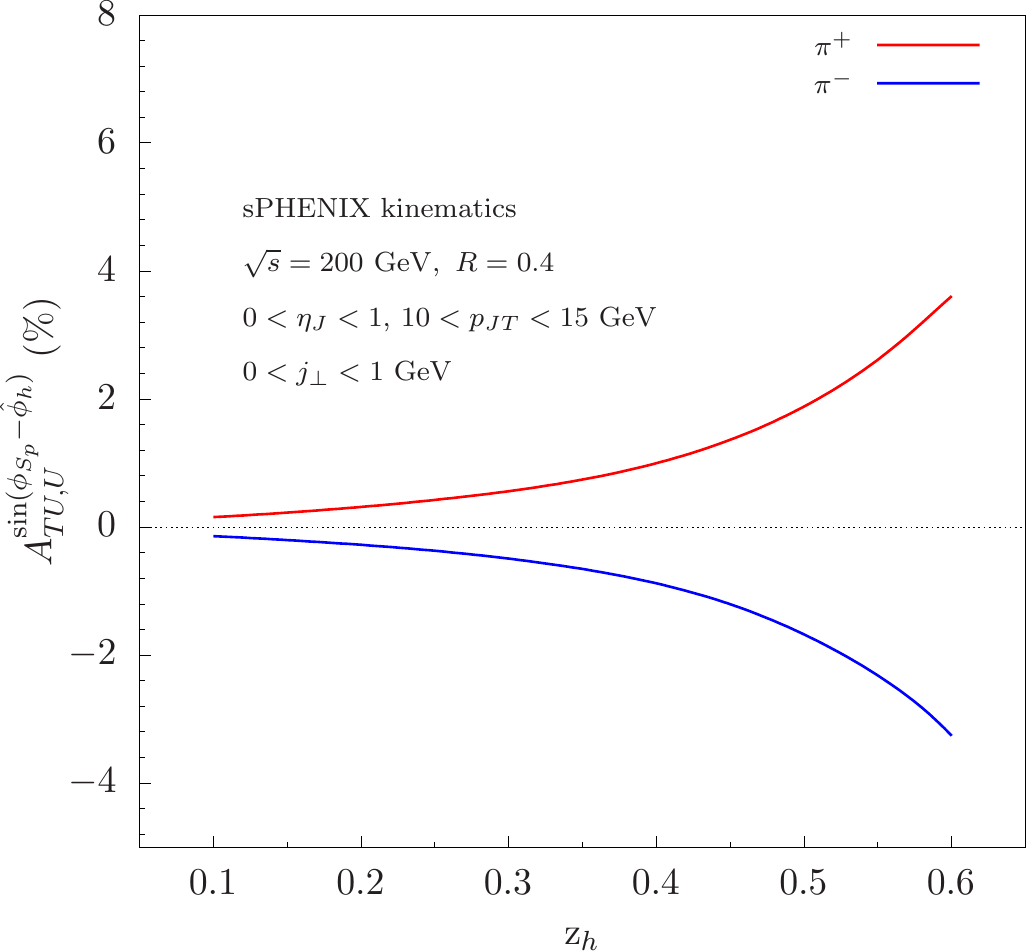} 
\caption{Left: Predictions for the transverse polarization $P_\Lambda$ of $\Lambda$ hyperons inside the jets in unpolarized $pp$ collisions. Right: Predictions for the Collins asymmetry $A^{\sin(\phi_{S_p}-\hat{\phi}_h)}_{TU,U}$ for $\pi^\pm$ inside the jet produced in $p^\uparrow p$ collisions, with red (blue) curve are for $\pi^+$ ($\pi^-$) production.}
\label{fig:PL}
\end{figure}

The left panel of Fig.~\ref{fig:PL} presents predictions for the $\Lambda$ transverse polarization, $P_\Lambda$, for $\Lambda$ hyperons inside the jets that are produced in unpolarized $p$+$p$ collisions at the center-of-mass (CM) energy $\sqrt{s}=200~$GeV, $p+p\to (\mathrm{jet}\,\Lambda^\uparrow)+ X$. Jets are constructed using the anti-$k_T$ jet algorithm with the radius $R = 0.4$ in the rapidity region $0 < \eta_J < 1$.  In such a collision, the transverse momentum ${\bf j}_\perp$ of the $\Lambda$ with respect to the jet axis and the transverse spin ${\bf S}_{h\perp}$ of the $\Lambda$ particle correlate with each other, generating a $\sin(\hat{\phi}_h - \hat{\phi}_{S_h})$ azimuthal dependence. The asymmetry is about $1-2\%$ and is a very promising measurement at sPHENIX. 

The right panel of Fig.~\ref{fig:PL} presents predictions of the Collins asymmetry $A^{\sin(\phi_{S_p}-\hat{\phi}_h)}_{TU,U}$ for unpolarized $\pi^\pm$ production inside the jets in transversely polarized $p^\uparrow$+$p$ collisions, $p^\uparrow+p\to (\mathrm{jet}\,\pi^\pm)+ X$. In this process, the transverse momentum ${\bf j}_\perp$ of $\pi^{\pm}$ with respect to the jet axis and the transverse spin ${\bf S}_{p}$ of the initial proton correlate with each other, generating a $\sin(\phi_{S_p}-\hat{\phi}_h)$ azimuthal dependence~\cite{Kang:2017btw,Yuan:2007nd}. Such an azimuthal asymmetry is referred to as Collins asymmetry and is sensitive to the quark transversity distribution in the transversely polarized proton, and the Collins fragmentation function for $\pi^{\pm}$. The asymmetry is positive for $\pi^+$ and negative for $\pi^-$ inside the jet and increases as $z_h$ increases. The magnitude of the asymmetry is about $4\%$, making measurements of this nature a good match to the expected large-statistics dataset of sPHENIX.


\section{Conclusion}

This manuscript collects physics predictions presented at a RIKEN-BNL Research Center (RBRC) workshop in July 2022 for the scientific program of sPHENIX, a next-generation collider detector at the Relativistic Heavy Ion Collider designed for a broad set of jet and heavy-flavor probes of the Quark-Gluon Plasma created in heavy ion collisions. 
sPHENIX is expected to begin commissioning and first data-taking in 2023, with a proposed running plan to enable precision measurements of reconstructed jet quenching, heavy flavor and quarkonia, cold QCD, and bulk physics. 
The detector features several experimental capabilities new to heavy-ion program at RHIC, including $b$-jet tagging, detailed jet sub-structure with very high statistics, and the potential observation of the $\Upsilon(3S)$ state in Au+Au collisions.
The contributions in this manuscript highlight compelling aspects of the physics program and, when possible, provide initial predictions in advance of sPHENIX data-taking to guide the development of the program. 
Confronting these predictions with data will be an important step towards elucidating the nature of the QGP created in heavy-ion collisions and completing the scientific mission of the RHIC facility.

\section*{Acknowledgements}

The authors gratefully acknowledge all workshop attendees for the fruitful scientific discussion, as well as Brookhaven National Laboratory (BNL) and the RIKEN-BNL Research Center (RBRC) for financial and logistical support.

Research support for individual authors is further acknowledged below. 
Funding from the U.S. Department of Energy (DOE) Office of Science has been provided under grants DE-SC0012704 (Y.H.), DE-FG02-03ER41244 (D.V.P.), DE-SC0013470 (M.S.), and DE-AC02-05CH11231 (X-N.W.). 
Funding from the National Science Foundation (NSF) has been provided under grants 1848162 (M.C.), PHY-1945471 (Z.K. and F.Z.), 2111046 (A.M.S.), and OAC-2004571 (R.E., A.K., and X-N.W.).
The work of M.D. is supported by the European Research Council, grant ERC-2016-COG: 725741, and by the Ministry of Science and Technological Development of the Republic of Serbia.
The work of M.\'{A}.E is supported by European Research Council project ERC-2018-ADG-835105 YoctoLHC, by the Maria de Maetzu excellence program under projects CEX2020-001035-M and CEX2019-000918-M, the Spanish Research State Agency under projects PID2020-119632GB-I00 and PID2019-105614GB-C21, the Xunta de Galicia (Centro singular de investigaci\'on de Galicia accreditation 2019-2022; European Union ERDF).
The work of G.G. is supported by the Deutsche Forschungsgemeinschaft (DFG, German Research Foundation) under Germany’s Excellence Strategy EXC 2181/1 - 390900948 (the Heidelberg STRUCTURES Excellence Cluster), SFB 1225 (ISOQUANT) and FL 736/3-1.
The work of J.H. is supported in part by the GLUODYNAMICS project funded by the ``P2IO LabEx (ANR-10-LABX-0038)'' in the framework ``Investissements d’Avenir'' (ANR-11-IDEX-0003-01) managed by the Agence Nationale de la Recherche (ANR), France.
The work of W.K. is supported by the US Department of Energy through the Office of Nuclear Physics and the LDRD program at Los Alamos National Laboratory
The work of D.P. is supported by funding from the European Union’s Horizon 2020 research and innovation program under the Marie Sklodowska-Curie grant agreement No. 754496. 
The work of D.V.P. is supported by funding from the Research Corporation for Scientific Advancement (RCSA).
The work of I.V. is supported in part by the LDRD program at LANL and by the U.S. Department of Energy under Contract No. 89233218CNA000001.
The work of Z.Y. is supported by the NSFC under grant Nos. 11935007, 11861131009 and 11890714.



\clearpage

\bibliographystyle{elsarticle-num} 
\bibliography{sphenix_workshop}





\end{document}